\documentclass[nofootinbib,prd,twocolumn,showpacs,showkeys,preprintnumbers]{revtex4-1}
\usepackage{hyperref,amssymb,amsmath,mathrsfs,bm,graphicx,comment,multirow}
\usepackage{array}

\usepackage[dvipsnames]{xcolor}

\usepackage[maxfloats=256]{morefloats}
\begin{document}

\title {Traversable wormholes with multiple unstable critical curves.}

\author{A. Rueda}
\email{alerue02@ucm.es}
\affiliation{Departamento de F\'isica Te\'orica,
	Universidad Complutense de Madrid, E-28040 Madrid, Spain.\\}

\author{E. Contreras }
\email{ernesto.contreras@ua.es}
\affiliation{Departamento de F\'{\i}sica Aplicada, Universidad de Alicante, Campus de San Vicente del Raspeig, E-03690 Alicante, Spain.\\}

\begin{abstract}
The number and position of unstable critical curves, as well as the nature of the accretion disk around compact objects, play a fundamental role in their optical appearance. Identifying differences in the optical spectrum of various observed compact objects can help classify them as black holes or black hole mimickers, such as traversable wormholes. Although multiple unstable critical curves have been reported to appear in asymmetric traversable wormholes, in this work we construct symmetric traversable wormholes with multiple unstable critical curves. We propose a general rational redshift function that allows us to trace the number of critical points of the effective potential and determine their nature as maxima or minima. The ray tracing method is used to study the trajectories of massless particles, particularly their behavior near the unstable critical points. Finally, a thin accretion disk model is implemented to analyze the optical appearance of the solution. 
\end{abstract}


\maketitle

\section{Introduction}
In recent years, astrophysical observations have made remarkable progress, particularly in the study of strong-field regimes of general relativity. The detection of gravitational waves by the LIGO/VIRGO collaboration \cite{LIGOScientific:2016aoc,LIGOScientific:2016sjg,LIGOScientific:2017bnn,LIGOScientific:2017ycc,LIGOScientific:2017vox} and the first image of a black hole (BH) shadow provided by the Event Horizon Telescope \cite{EventHorizonTelescope:2019dse,EventHorizonTelescope:2019uob,EventHorizonTelescope:2019jan,EventHorizonTelescope:2019ths,EventHorizonTelescope:2019pgp,EventHorizonTelescope:2019ggy} are milestones that strongly support the predictions of general relativity. However, these observations also open the door to exploring new physics, as distinguishing black holes from their potential mimickers—such as traversable wormholes—requires examining subtle observational details. For instance, while the shadows of M87* and Sgr A* align with theoretical expectations for black holes, alternative compact objects capable of producing similar shadows cannot yet be ruled out \cite{Guerrero:2021pxt}. 

A promising way to characterize compact objects is through their optical appearance, determined by the emission spectrum of the accretion disk surrounding them. This spectrum
typically consists of a series of emission patterns or rings of light, whose structure depends on the number of unstable critical curves—null trajectories where the energy coincides with the maximum of the effective potential. For instance, in the Schwarzschild case, there is a single critical curve, and its optical appearance comprises a specific pattern of light rings depending on the emission model \cite{Gralla:2019xty}. This dependence on critical curves provides an opportunity to differentiate between black holes and their mimickers, as the number and configuration of rings of light arising from multiple maxima in the effective potential could result in observable differences. These differences could, in principle, be observed using radio interferometry with extremely high angular resolution, a capability not yet fully realized. 

Wormhole solutions with double critical curves have been previously studied, particularly in asymmetric configurations of the effective potential with respect to the wormhole's throat
\cite{Morris:1988cz,Morris:1988tu,Visser:1995cc,Visser:2003yf,Alcubierre:2017pqm,Lobo:2005us,Blazquez-Salcedo:2021udn,Bambi:2021qfo,Capozziello:2020zbx,Blazquez-Salcedo:2020czn,Berry:2020tky,Maldacena:2020sxe,Guerrero:2021pxt,Wielgus:2020uqz,Blazquez-Salcedo:2020nsa}. More recently, it was rigorously demonstrated that traversable wormholes always possess at least one critical curve, whether the wormhole is symmetric or asymmetric \cite{Xavier:2024iwr}. The study further concludes that asymmetry influences the position of critical curves, while symmetric wormholes invariably have one located at the throat. Based on these results, in this work, we construct symmetric traversable wormholes with multiple unstable critical curves. A key feature of our model is the inclusion of a non-zero redshift function, $\Phi(r) \neq 0$, which has significant implications for the geometry and observables of the wormhole spacetime. While a zero redshift function eliminates radial tidal forces, it also limits the phenomenological richness of the solution, particularly in terms of the effective potential and the existence of multiple unstable critical curves. In contrast, a non-zero $\Phi(r)$ introduces radial energy gradients that allow for the appearance of additional critical curves, which are essential for producing distinct optical signatures in the wormhole's observable properties. This choice is not arbitrary; instead, it reflects realistic physical scenarios where gravitational redshift affects photon trajectories and emission patterns. It should also be emphasized that this work does not aim to analyze the magnitude of the tidal forces in our wormhole model but rather to propose a redshift function that enables the emergence of additional critical curves. Notably, the shape function $b(r)$, which primarily governs the throat geometry, plays no significant role in generating multiple critical curves, highlighting the centrality of the redshift function in this context. To complement the work, we analyze the behavior of null geodesics near these curves using the ray-tracing method and investigate the optical appearance of each solution by modeling a thin accretion disk. It is worth mentioning that in the literature, the term light ring is used inconsistently. Some works define it as unstable critical curves, others as any critical curve (whether stable or unstable), and still others use it to describe the circular patterns in the optical spectrum of accretion disks. To avoid ambiguity, we refrain from using the term light ring and instead refer explicitly to unstable critical curves, maxima/minima of the potential, or optical features depending on the context.

This work is organized as follows. In the next section, we introduce a brief review on traversable wormholes physics. Next, in section \ref{geodesic} we explore the main aspects associated to the motion of massless particles around a wormhole geometry emphasizing the motion near the critical curves. Then,  in section  \ref{solution} we construct our model of wormholes with $n$ critical curves, and develop the cases blue $n = 0,1$ and $2$ in detail. In section \ref{nullgeodesics} we employ the ray-tracing method and show the trajectories followed by null geodesic for different values of the parameters involved. In section \ref{thin} we develop a model of a thin accretion disc around our wormhole geometry and obtain the observed intensity and its optical appearance. Finally, the last section is devoted to the conclusions of the work.

\section{Traversable wormholes}\label{TW}
Let us consider the spherically symmetric line element \cite{Morris:1988cz,Morris:1988tu,Visser:1995cc,Visser:2003yf}
\begin{eqnarray}\label{metric}
ds^{2}=-e^{2\Phi} dt^2 +d r^{2}/(1-b/r)+r^{2}(d\theta^{2}+\sin^{2}\theta d\phi^{2}),\nonumber\\
\end{eqnarray}
with $\Phi=\Phi(r)$ and $b=b(r)$ the redshift and shape functions respectively. 
Assuming that (\ref{metric}) is a solution of Einstein's equations
\begin{eqnarray}\label{EFE}
R_{\mu\nu}-\frac{1}{2}g_{\mu\nu}R=\kappa T_{\mu\nu},
\end{eqnarray}
with $\kappa=8\pi G/c^{2}$ \footnote{In this work we shall assume $c=G=1$.}, sourced by 
$T^{\mu}_{\nu}=diag(-\rho,p_{r},p_{t},p_{t})$
we arrive at
\begin{eqnarray}
\rho&=&\frac{1}{8\pi}\frac{b'}{r^{2}}\label{rho}\\
p_{r}&=&-\frac{1}{8\pi}\left[\frac{b}{r^{3}}-2\left(1-\frac{b}{r}\right)\frac{\phi'}{r}\right]\label{pr}\\
p_{t}&=&\frac{1}{8\pi}\left(1-\frac{b}{r}\right)
\bigg[\phi''+(\phi')^{2}-\frac{b'r-b}{2r^{2}(1-b/r)}\phi'\nonumber\label{pt}\\
&&-\frac{b' r-b}{2r^{3}(1-b/r)}+\frac{\phi'}{r}\bigg].
\end{eqnarray}
%As there is not horizon, $g_{tt}$ must be a non vanishing function to avoid the existence of a infinite redshift surface, then $\phi$ must be finite everywhere.\\
%The information about the throat of the wormhole is encoded in its shape. First, given that our solution is spherically symmetric, 

In what follows we shall describe the main aspects of a traversable wormhole by its embedding in  the three dimensional Euclidean space. First, note that as our solution is spherically symmetric, 
we can consider $\theta=\pi/2$ without loss of generality. Now, considering a fixed time, $t=constant$, the line element reads
\begin{eqnarray}\label{emb1}
ds^{2}=\frac{dr^{2}}{1-b/r}+r^{2}d\phi^{2}.
\end{eqnarray}
The surface described by (\ref{emb1}) can be embedded in $\mathbf{R}^{3}$ where the metric in cylindrical coordinates $(r,\phi,z)$ reads
\begin{eqnarray}
ds^{2}=dz^{2}+dr^{2}+r^{2}d\phi^{2}.
\end{eqnarray}
Next, as $z$ is a function of the radial coordinate we have
\begin{eqnarray}
dz=\frac{dz}{dr}dr,
\end{eqnarray}
from where
\begin{eqnarray}\label{emb2}
ds^{2}=\left[1+\left(\frac{dz}{dr}\right)^{2}\right]dr^{2}
+r^{2}d\phi^{2}.
\end{eqnarray}
Finally, from (\ref{emb1}) and (\ref{emb2}) we obtain
\begin{eqnarray}\label{emb3}
\frac{dz}{dr}=\pm 
\left(\frac{r}{b}-1\right)^{-1/2},
\end{eqnarray}
where is clear that the condition $b<r$
must be satisfied. At this point some comments are in order. First, the  traversable wormhole geometry must be endowed with a throat, namely, a minimum radius $r_{0}$  where $dz/dr\to\infty$, so that $b(r=r_{0})=b_{0}=r_{0}$. Second, we demand that the solution is asymptotically flat which implies both, $b/r\to0$ (from where $dz/dr\to0$) and $\Phi\to0$ as $r\to\infty$.  Third, as
as the conditions 
\begin{eqnarray}
&&\lim\limits_{r\to r_{0}}\frac{dz}{dr}\to\infty\label{c1}\\
&&\lim\limits_{r\to\infty}\frac{dz}{dr}=0\label{c2},
\end{eqnarray} 
must be satisfied, the smoothness of the geometry is ensured whenever the embedding surface flares out at or near the throat, namely  
\begin{eqnarray}
\frac{d^{2}r}{dz^{2}}>0,
\end{eqnarray}
from where
\begin{eqnarray}\label{foc}
\frac{b-b'r}{2b^{2}}>0,
\end{eqnarray}
which corresponds to the 
flaring--out condition. 

It is worth mentioning that, the flaring out condition (\ref{foc}) leads to the violation of the null energy condition (NEC) as we shall see in what follows. Let us define the quantity
\begin{eqnarray}
\xi=-\frac{p_{r}+\rho}{|\rho|}=\frac{b/r-b'-2(r-b)\phi'}{|b'|},
\end{eqnarray}
which can be written as
\begin{eqnarray}
\xi=\frac{2b^{2}}{r|b'|}\frac{d^{2}r}{dz^{2}}
-2(r-b)\frac{\phi'}{|b'|}
\end{eqnarray}
Now, as $(r-b)\to0$ at the throat, we have 
\begin{eqnarray}
\xi=\frac{2b^{2}}{r|b'|}\frac{d^{2}r}{dz^{2}}>0
\end{eqnarray}
so that
\begin{eqnarray}
\xi=-\frac{p_{r}+\rho}{|\rho|}>0.
\end{eqnarray}
Note that if $\rho>0$ the above condition implies $p_{r}<0$ which entails that $T^{1}_{1}$ should be interpreted as a tension. Furthermore, if we define $\tau=-p_{r}$ the flaring out condition leads to
\begin{eqnarray}
\tau-\rho>0,
\end{eqnarray}
which implies that, the throat tension must exceed the total energy density, which, as stated before, violates the NEC and therefore requires exotic matter to support the wormhole. It is worth mentioning that the requiring of exotic matter could translate in introducing classical phantom field in the matter sector. Nevertheless, recent studies have shown that such phantom fields may not be mandatory (see \cite{Konoplya:2021hsm,Blazquez-Salcedo:2020czn}, for example).

\section{Geodesic Equations and Critical Curves}\label{geodesic}

The geodesic equations for a static and spherically symmetric spacetime, with constant $\theta$, are

\begin{align}
&\dot{t} = E e^{-2\Phi(r)},\\
&\dot{\phi} = \frac{L}{r^2 \sin^2\theta},\\
&\dot{r}^2 = \left( 1 - \frac{b(r)}{r} \right) \left( E^2 e^{-2\Phi(r)} - \epsilon - \frac{L^2}{r^2 \sin^2\theta} \right),\label{rdot}
\end{align}
where $L$ and $E$ are the angular momentum and energy of the particle, respectively. Additionally, $\dot{r}$ represents the derivative of $r$ with respect to an affine parameter, and $\epsilon = 1, 0$ corresponds to timelike and null geodesics, respectively.

It is convenient to introduce the tortoise coordinate since it allows us to connect both universes without coordinate divergences. The tortoise coordinate $x$ is defined as
\begin{eqnarray}\label{tort}
\frac{dx}{dr} = \frac{1}{\sqrt{1 - b(r)/r}}.
\end{eqnarray}
The equation \eqref{rdot} becomes
\begin{align}\label{eqgeo1}
\dot{x}^2 = e^{-2\Phi}(E^2 - V_\pm^2(x,\theta)),
\end{align}
where $V_\pm(x,\theta)$ is the effective potential that depends on $x$ and $\theta$, defined as
\begin{equation}\label{effpot}
V_{\pm}(x,\theta) = \pm \sqrt{\epsilon + L^2 \frac{e^{2\Phi(x)}}{r(x)^2 \sin^2\theta}}.
\end{equation}

For null geodesics, we define the impact parameter $\beta = L/E$, and equation \eqref{eqgeo1} becomes

\begin{align}\label{eqgeonull}
\frac{\dot{x}^2}{L^2} = \frac{1}{\beta^2} - \frac{e^{2\Phi(x)}}{r(x)^2 \sin^2\theta},
\end{align}
and the effective potential \eqref{effpot} is
\begin{equation}
V_{\pm}(x,\theta) = \pm \frac{e^{\Phi(x)}}{r(x) \sin\theta}.
\end{equation}
From \eqref{eqgeonull}, we can define, for each impact parameter $\beta$, the nearest point  from the throat that the photon's trajectory reaches, $x_{min}$. This point satisfies $\dot{x}|_{x_{min}} = 0$, which can be calculated by solving
\begin{equation}\label{rmin}
\frac{1}{\beta^2} = \frac{e^{2\Phi(x_{min})}}{r(x_{min})^2}
\end{equation}
for $x_{min}$, where we have set $\theta=\pi/2$ for simplicity, without loss of generality, as the solution is spherically symmetric.
The critical curves are defined as circular null geodesics and the impact parameter of a photon following this trajectory are known as the critical impact parameter $\beta_c$. The conditions for the effective potential for critical curves are
\begin{equation}\label{impact}
\frac{1}{\beta_c^2} = V_\pm(x_c), \:\; \frac{dV_\pm}{dx}\bigg|_{x_c} = 0, \:\; 
\end{equation}
where $x_c$ is the radius of the circular orbit. Although it is clear that $ x_{c} $ can be located anywhere, it can be demonstrated that for a traversable wormhole with a null redshift function ($ \Phi = 0 $), there exists an unique unstable critical curve located at the throat. To demonstrate that, let us consider a spherically symmetric spacetime with the parametrization \eqref{metric}, and assume that there exists a $r_0$ such that $b(r_0) = r_0$. Now, we can explicitly calculate $dV/dx$ at $x(r_0) = 0$ to see if there is a critical curve at the throat (for $\theta=\pi/2$), namely
\begin{equation}
    \begin{split}
        \frac{d V}{dx} =  e^{\Phi(x)} \frac{dr}{dx} \left( \frac{d\Phi}{dr} - \frac{1}{r} \right).
    \end{split}
\end{equation}
Using the definition of the coordinate $x$ \eqref{tort} we obtain
\begin{equation}\label{derivada del pot}
    \begin{split}
        \frac{d V}{dx} =  e^{\Phi(x)} \sqrt{1 - b(x)/r(x)} \left( \frac{d\Phi}{dr} - \frac{1}{r} \right)
    \end{split}
\end{equation}
from where is clear that
at the throat $x(r_0) = 0$, we get
\begin{equation}
    \begin{split}
        \frac{d V}{dx}\bigg|_{x(r_0)} = 0
    \end{split}
\end{equation}
This implies that,   as we stated before, in the particular case where $\Phi = 0$, there exists a unique critical curve at $x(r_0) = 0$, which is unstable as can be easily verified. In this regard, the role of the redshift function $\Phi(r)$ is central to the construction of wormhole geometries with multiple critical curves. Although a zero redshift function has the advantage of eliminating radial tidal forces, a non-zero $\Phi(r)$ offers observational advantages by enabling richer and more complex optical signatures as we will see in the next sections.

\section{Travsersable Wormhole 
Model with multiple Critical Curves}\label{solution}
As we have already demonstrated that $\Phi = 0$ leads to a solution with a single critical curve, we aim to find the conditions that must be met to ensure $dV/dx=0$ when $\Phi \ne 0$ for some point $r_{c}\ne r_{0}$. Then, we must classify the unstable critical curves which are associated to maximum of the effective potential, namely those leading to $d^{2}V/dx^{2}<0$. 

By using \eqref{derivada del pot}, we find that the condition for $r_{c}$ to exist is given by
\begin{eqnarray}\label{condition pot 1}
    \frac{d\Phi}{dr} \bigg|_{r_c} = \frac{1}{r_c}.
\end{eqnarray}

Now, computing the second derivative of the effective potential, for null geodesics we arrive at
%\begin{eqnarray}\label{condition pot 2}
%\frac{d^2 V_\pm}{dx^2} = \frac{dr}{dx} \frac{d}{dr} \left( e^{\Phi} \frac{dr}{dx}\right) \left( \frac{d\Phi}{dr} - \frac{1}{r} \right) + \left(\frac{dr}{dx}\right)^2 e^{\Phi} \left(\frac{d^2 \Phi}{dr^2} + \frac{1}{r^2} \right).
%\end{eqnarray}
%Evaluating \eqref{condition pot 2} at a critical point of $V_\pm$, $r_c \neq r_0$ (satisfying \eqref{condition pot 1}), we have

\begin{eqnarray}
 \frac{d^2 V_\pm}{dx^2}\bigg|_{r_c} = e^{\Phi(r_c)} \left( \frac{dr}{dx}\bigg|_{r_c}\right)^2 \left(\frac{d^2 \Phi}{dr^2}\bigg|_{r_c} + \frac{1}{r^2_c}\right),
\end{eqnarray}
form where is clear that 
the stability of the light orbit at $r_c$, depends on the sign of the term
\begin{eqnarray}\label{segundad}
    \frac{d^2 \Phi}{dr^2}\bigg|_{r_c} + \frac{1}{r^2_c}.
\end{eqnarray}

To ensure the physical relevance of our model, we propose a general rational form for the redshift function, which provides both flexibility and control over the number and location of critical points in the effective potential, namley
\begin{eqnarray}\label{phin}
    \Phi_n(r) = \frac{a_n r^n + a_{n-1}r^{n-1} + \dots + a_1 r + a_0}{r^{n+1}}.
\end{eqnarray}
Note that by assuming that there are $n+1$ points, $r_1, r_2, \dots, r_{n+1}$, each  one satisfying \eqref{condition pot 1}, the resulting $n+1$ equations, leads to an expression of the coefficients $a_i$ in terms of the positions of the critical points, $r_{i}$ given by
\begin{equation}\label{formula coef}
    \begin{split}
        a_n &= -(r_1 + \dots + r_{n+1}) \\
        a_{n-k} &= \frac{(-1)^{k+1}}{k+1} \sum_{i_1=1}^{n-k-1}\sum_{i_2=i_2+1}^{n-k} \dots \sum_{i_{k+1}=i_k+1}^{n+1} r_{i_1}r_{i_2}\dots r_{i_{k+1}} \\
        a_0 &= \frac{(-1)^{n+1}}{n+1} r_1 r_2 \dots r_{n+1} \\
    \end{split}
\end{equation}
with $k=0, 1, \dots, n-1$. For example, for the cases $n = 0,1,2$, we have
%\begin{eqnarray}
 %   \Phi_0(r) = \frac{a_0}{r}
%\end{eqnarray}
%Using the formula for the coefficients \eqref{formula coef}, we obtain
\begin{eqnarray}
\Phi_0(r) &=& -\frac{r_1}{r}\\
\Phi_1(r) &=& \frac{-(r_1+r_2)r + \frac{1}{2}r_1r_2}{r^2}\\
    \Phi_2(r) &=& \frac{a_2r^2 + a_1r + a_0}{r^3}
\end{eqnarray}
where
\begin{equation}
\begin{split}
    a_2 &= -(r_1 + r_2 + r_3) \\
    a_1 &= \frac{1}{2}(r_1r_2 + r_1r_3 + r_2r_3) \\
    a_0 &= -\frac{1}{3}r_1r_2r_3
\end{split}
\end{equation}

\begin{figure*}[hbt!] 
\centering
 \includegraphics[width=0.3\textwidth]{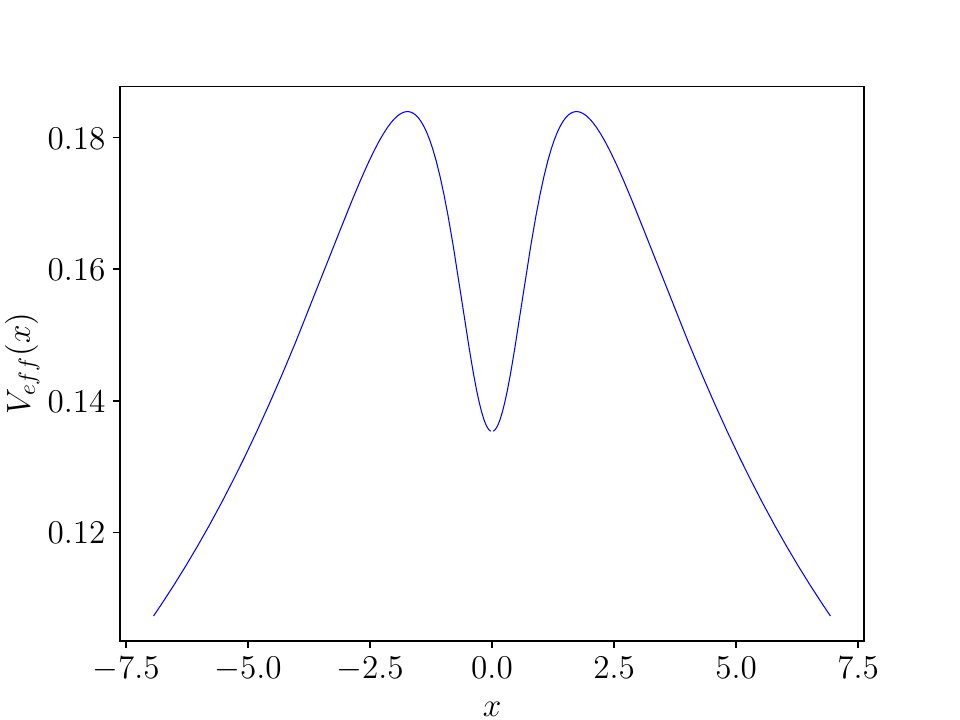}\
\includegraphics[width=0.3\textwidth]{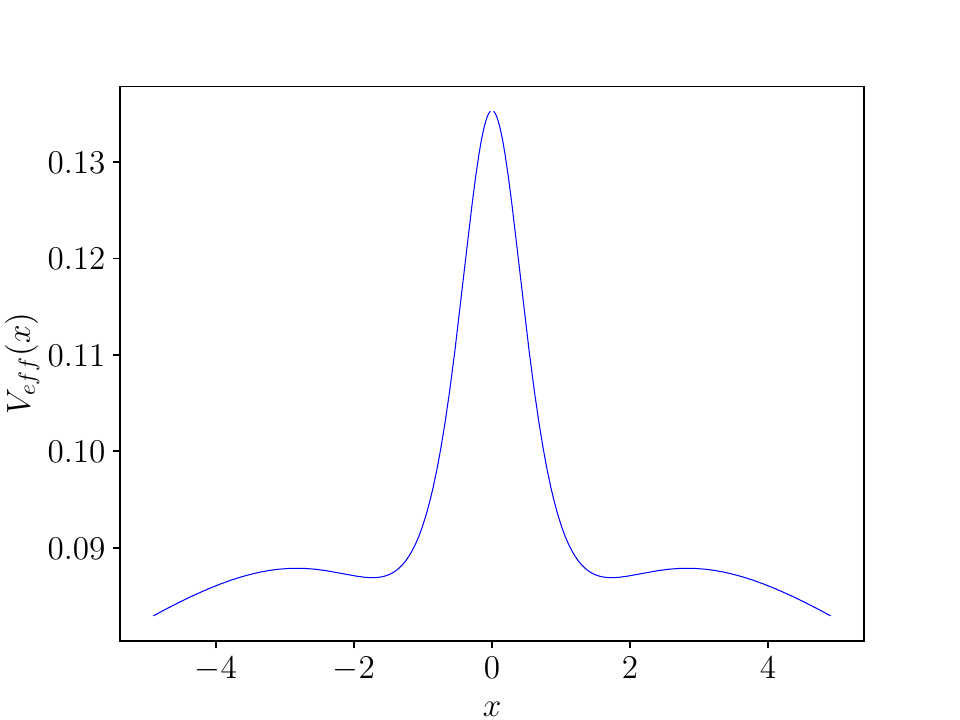}\
\includegraphics[width=0.3\textwidth]{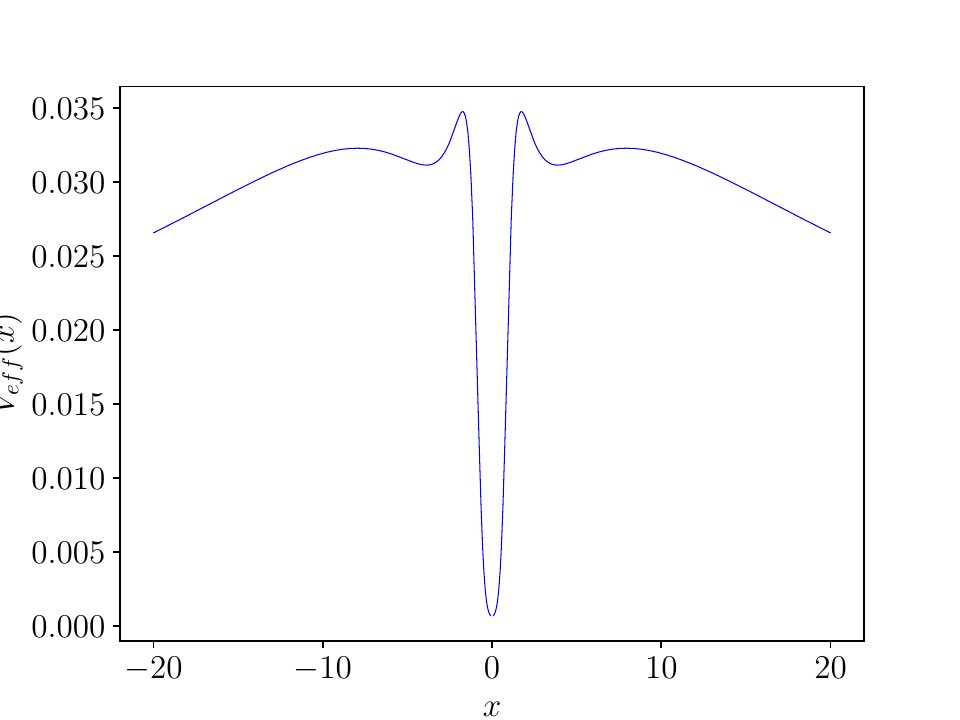}\
\includegraphics[width=0.3\textwidth]{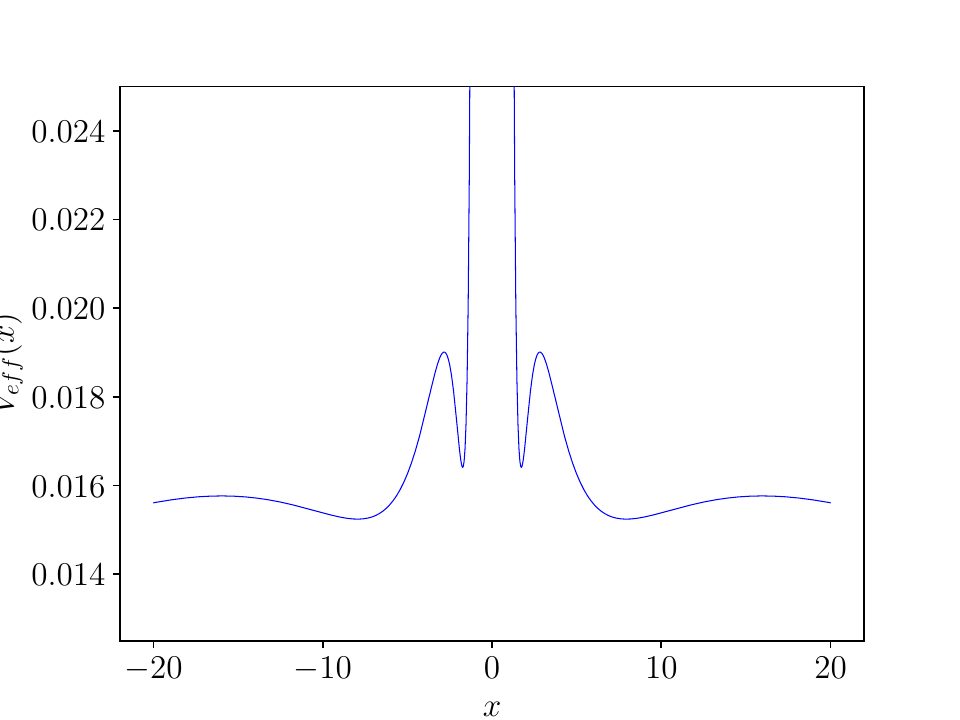}\

\caption{\label{pots1}
Effective potentials $V_n$ for $n = 0, 1, 2$ (top row) and $n = 3$ (bottom row). For even $n$ the potential  has a minimum at the throat, while for odd $n$ it has a maximum}.
\end{figure*}

We note that, each function $\Phi_n$ gives a potential $V_n$ with $n+1$ critical points. To analyze the sign of the second derivative of $V_n$ at the critical points, we assume $r_0 < r_1 < \dots < r_i < \dots < r_{n+1}$. Using \eqref{segundad}, we calculate the sign of $d^2V_n/dx^2$, obtaining
\begin{equation}\label{segundad2}
\begin{split}
    &\left. \frac{d^2 \Phi_n}{dr^2} + \frac{1}{r^2} \right|_{r_i} = \\
    & -\frac{1}{r_i^{n+2}}(r_i-r_1) \dots (r_i-r_{i-1}) (r_i-r_{i+1}) \dots (r_i-r_{n+1})
\end{split}
\end{equation}
where $i=1, \dots, n+1$. At this point some comments are in order. First, note that the term $(r_i - r_j)$ is either positive if $i > j$ or negative if $i < j$. Second, 
if we evaluate \eqref{segundad2} at $r_1$ (the smallest critical point distinct from $r_0$) we have that when $n$ is even, there are even numbers of terms in \eqref{segundad2}, and at $r_1$, all terms are negative, so $\left. \frac{d^2 \Phi_n}{dr^2} + \frac{1}{r^2} \right|_{r_1} < 0$. In this regard, the potential $V_n$ has a maximum at $r_1$, so $V_n$ has a minimum at $r_0$ and at $r_2$. Similarly, if $n$ is odd, $V_n$ has a minimum at $r_1$, so there is a maximum at $r_0$ and at $r_2$. Finally, the effective potential corresponding to null geodesics for the geometry with $\Phi_n$ \eqref{phin} is
\begin{equation}
    V_n(x) = \frac{e^{\Phi_n(x)}}{r(x)},
\end{equation}
which satisfies
\begin{equation}
    \dot{x}^2 = \frac{1}{\beta^2} - V_n^2(x).
\end{equation}
To explore the profile of the effective potential, we must propose a particular shape function. In this work, we consider $b = r_{0}^{2}/r $ without loss of generality. It is obvious that we could have used any other, but this one is the most convenient because it simplifies the numerical calculations that we will do later. The effective potentials for  $n = 0, 1, 2, 3$  are shown in Figure \ref{pots1}. The location of the critical points is summarized in Table \ref{table1}. It is worth noticing that, for the case  $n = 3$, the potential is very high, so most light rays would be unable to traverse the wormhole. It can be seen that, for even (odd) $n$, $V_n$ has a minimum (maximum) at the throat that coincides with our discussion above.
\\

\begin{table}[]
\begin{tabular}{|c|c|c|c|c|}
\hline
$n$ & $r_1$ & $r_2$ & $r_3$ & $r_4$ \\ \hline

0     & $2r_0$ &         &         &         \\ \hline

1     & $2r_0$ & $3r_0$ &         &         \\ \hline

2     & $2r_0$ & $3r_0$ & $8r_0$ &         \\ \hline

3     & $2r_0$ & $3r_0$ & $8r_0$ & $16r_0$ \\ \hline
\end{tabular}
\caption{Critical points $r_c$ for wormhole models with $n = 0, 1, 2, 3$. These points correspond to the radii where the effective potential $V_n(x)$ has maxima or minima \label{table1}}
\end{table}

\section{Ray-Tracing}\label{nullgeodesics}

To obtain the ray-tracing of the wormhole geometry we numerically integrate 
\begin{equation} \label{eqgeo}
    \frac{d\phi}{dr} = \pm \frac{1}{r^2\sqrt{(1-\frac{b(r)}{r})(\frac{e^{-2\Phi}}{\beta^2}-\frac{1}{r^2})}}
\end{equation}
with the sign  $+$ for outgoing geodesics and $-$ for ingoing geodesics.
For the $+$ sign for incoming orbits, from a numerical initial point at infinity to $ r_{min} $ \eqref{rmin}, which is the closest point to the throat along the light ray's path. Similarly, we numerically integrate \eqref{eqgeo} with the $ - $ sign for outgoing orbits, from $ r_{min} $ to infinity.

\begin{figure*}[hbt!]
\centering
 \includegraphics[width=0.4\textwidth]{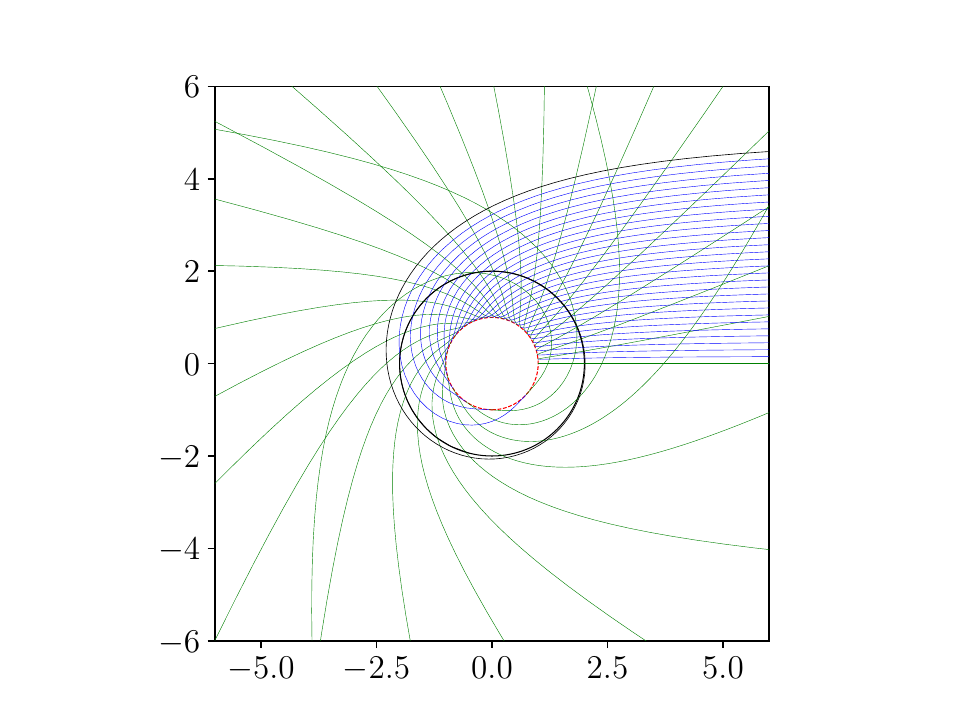}\
\includegraphics[width=0.4\textwidth]{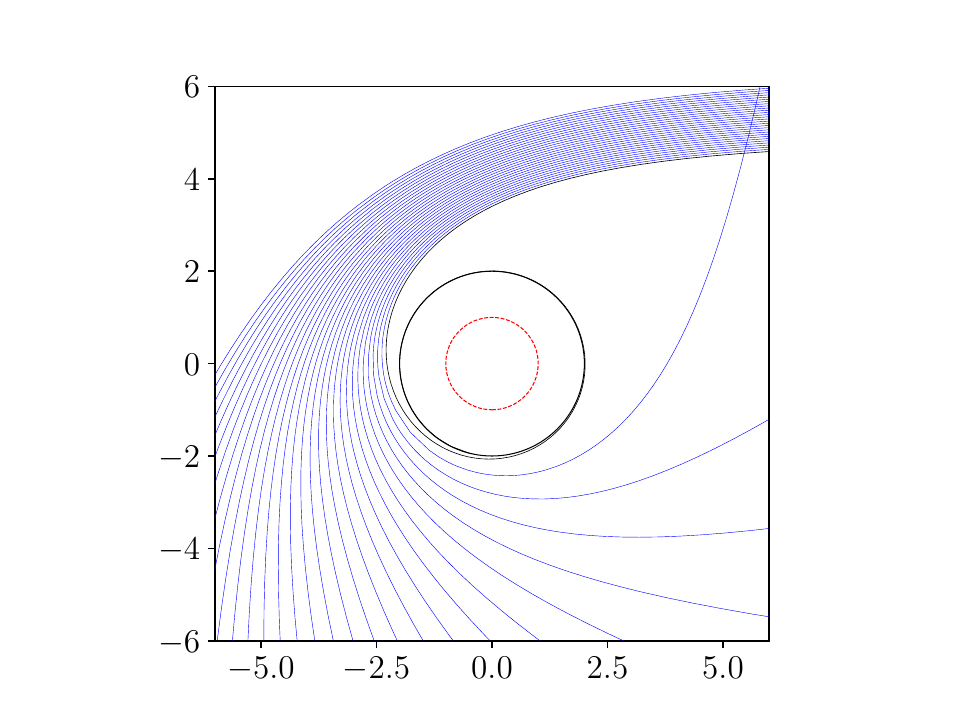}\

\caption{\label{rtn1}
Ray-Tracing for \
$n = 0$ . Left panel: 
The blue curves represent the trajectory of photons approaching to the throat
with impact parameter $\beta \in [0,\beta_{c})$ and the green curves represent the outgoing geodesic in the other asymptotically flat region (or another universe). Right panel: Trajectories of photons with impact parameter $ \beta \in [\beta_{c},\beta_{max} =10] $. In this case, the photons remain in the same universe. In both cases, the outer circle corresponds to the unstable critical point $ r_1$  while the red inner circle corresponds to the throat $r_0$. The trajectory associated to the unstable critical curve corresponds to the black curves which is formed at $r_1$}
\end{figure*}

\begin{figure*}[hbt!]
\centering
 \includegraphics[width=0.4\textwidth]{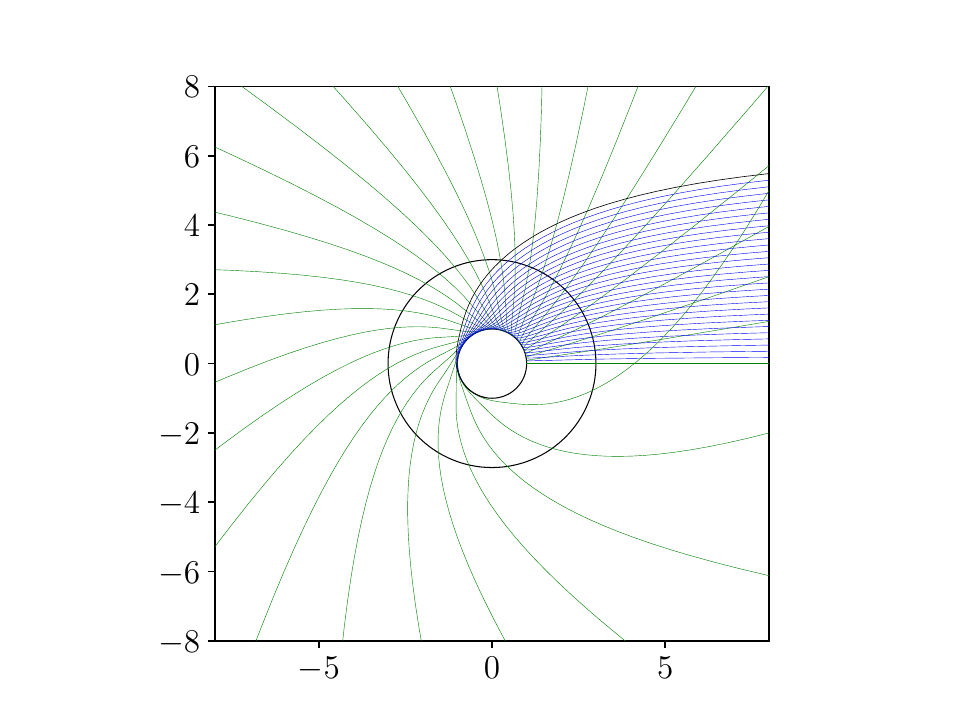}\
\includegraphics[width=0.4\textwidth]{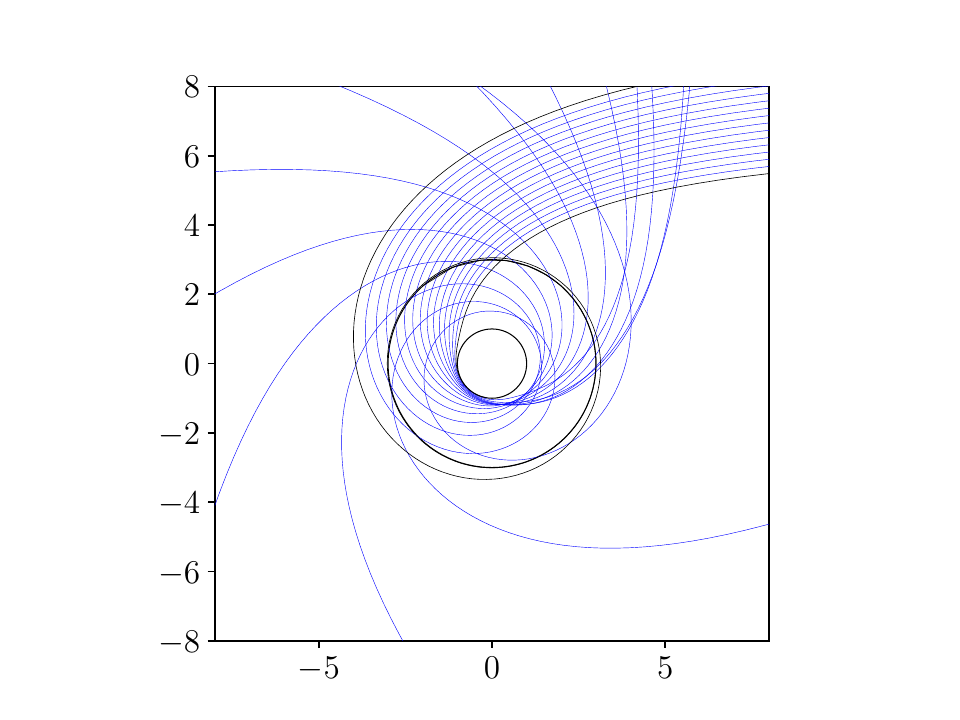}\
\includegraphics[width=0.4\textwidth]{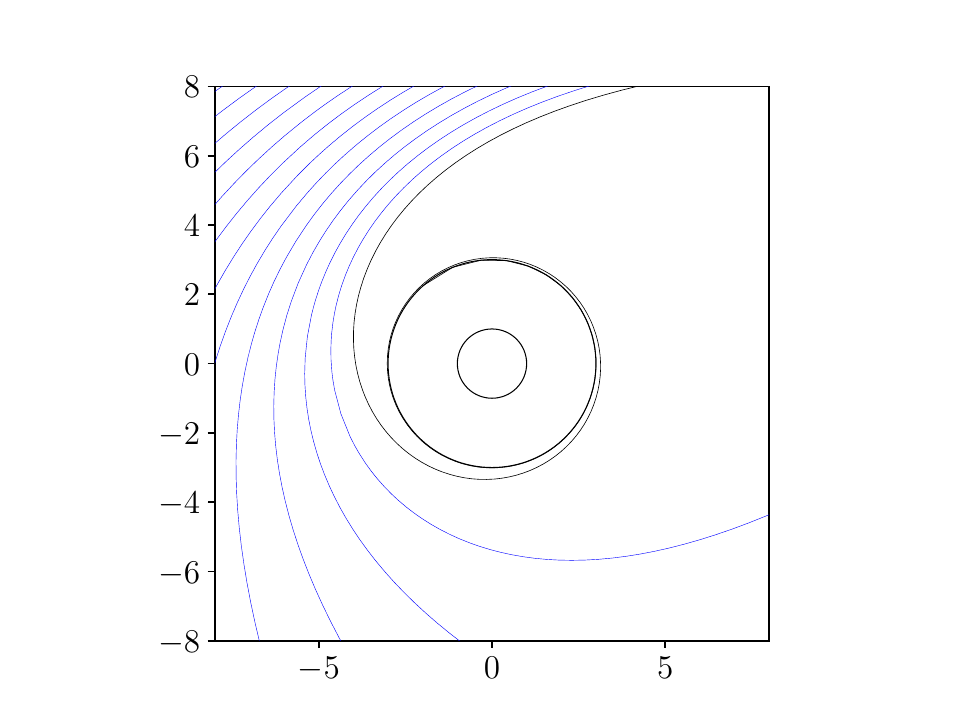}\
\caption{\label{rtn2}
Ray-Tracing for  
$n = 1$. First row, left panel: Trajectory of incoming photons represented with blue curves with impact parameter
$\beta\in [0,\beta_{c0})$. These photons traverse the throat and appear in the other asymptotic region or universe and are represented with green curves. The unstable critical curve appears at the throat (inner circle) and the trajectory associated to it is the black curve. First row, right panel: Trajectory of photons approaching to the throat with impact parameter $\beta \in [\beta_{c0},\beta_{c2})$. In this case, the whole photons remain in the same universe and there are two unstable critical curves, at the critical circles $r_0$ and $r_2=3r_0$. Second row: Photons with impact parameter $\beta \in [\beta_{c2},\beta_{max} =16]$. In this case the unstable critical curve appears only at $r_2$.}
\end{figure*}

\begin{figure*}[hbt!]
\centering
 \includegraphics[width=0.4\textwidth]{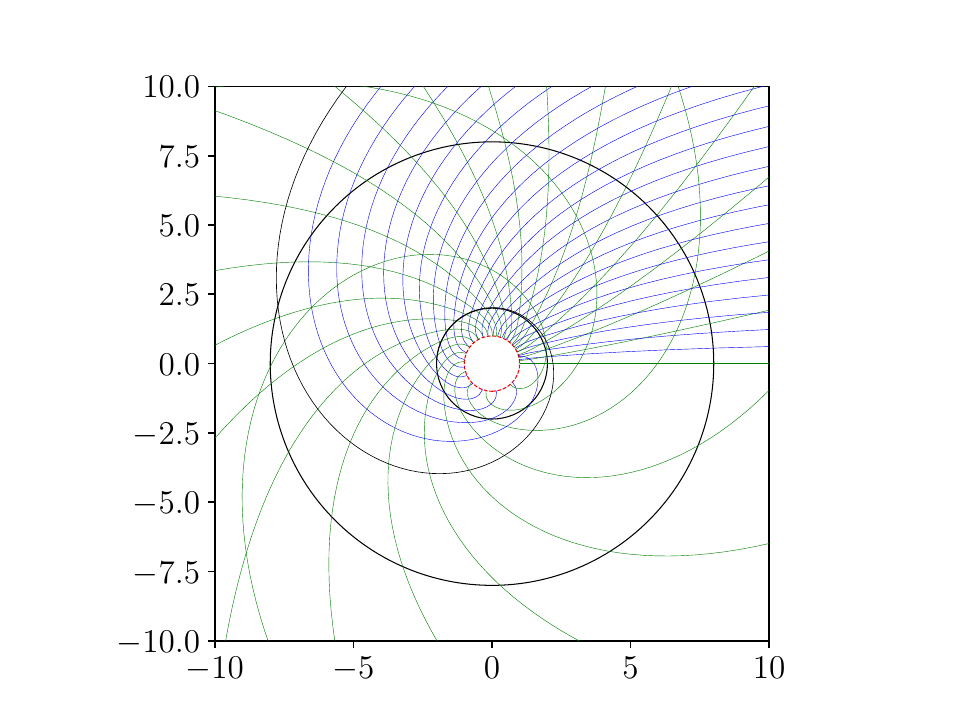}\
\includegraphics[width=0.4\textwidth]{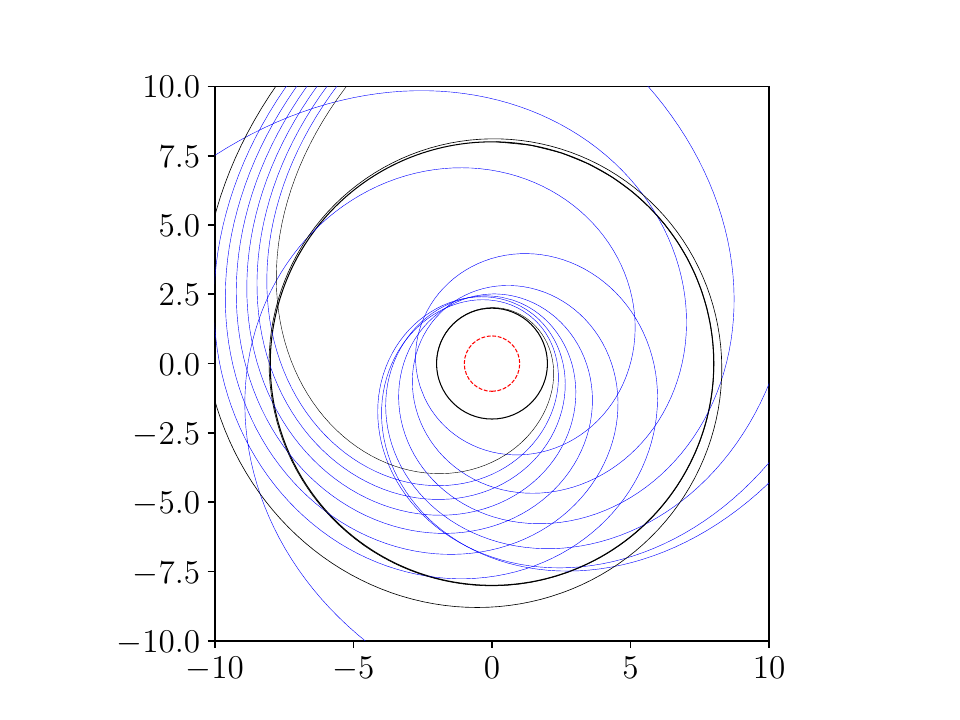}\
\includegraphics[width=0.4\textwidth]{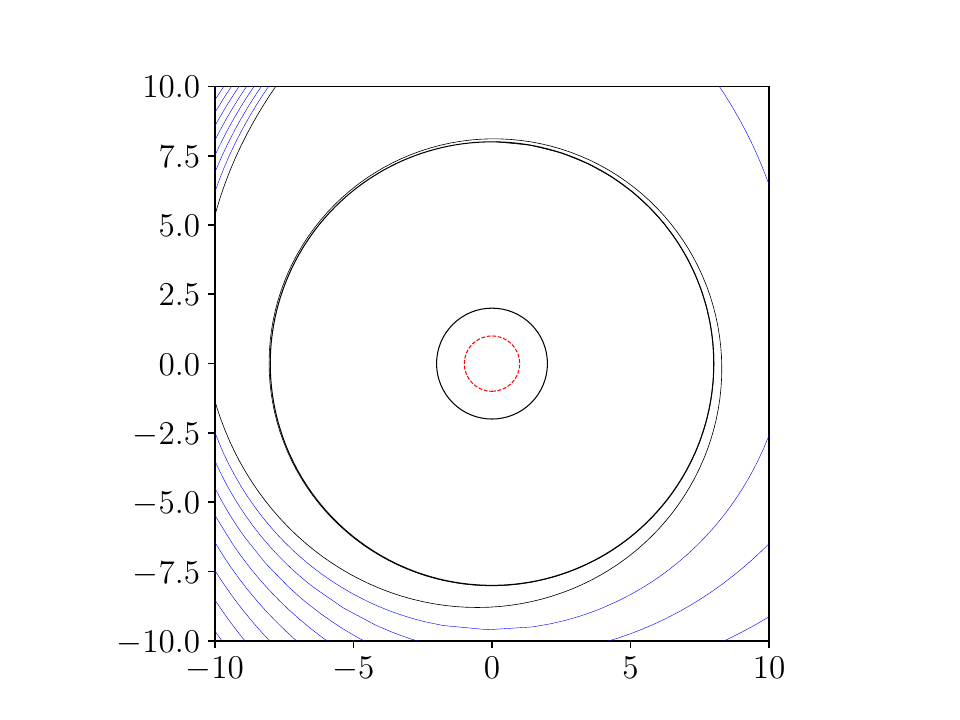}\

\caption{\label{rtn3}
Ray-Tracing for 
$ n = 2 $. First row, left panel: Incoming photons (blue curve) with impact parameter $\beta \in [0,\beta_{c1})$, outgoing photons in the other asymptotic region (green curves) and unstable critical curve (black curve). First row, right panel. All the photons with impact parameter $\beta \in [\beta_{c1},\beta_{c3})$ remain in the same universe. Now, there are two unstable critical curves at $r_1$ and $r_3$. Second row: Impact parameter
$ \beta \in [\beta_{c3},\beta_{max} = 50] $ There is a unstable critical curve at $r_3$. The black circles correspond to the critical curves at $r_1$ and $r_3$, the red inner
circle corresponds to the throat $r_0$.}
\end{figure*}
Taking $ r_i $ and $ \phi_i $ as the initial position, the incoming geodesic satisfies
\begin{equation}\label{geo1}
    \phi_{in} (r) =  \int_{r_i}^{r} \frac{dr'}{r'^2\sqrt{(1-\frac{b(r')}{r'})(\frac{e^{-2\Phi}}{\beta^2}-\frac{1}{r'^2})}}  + \phi_i,
\end{equation}
where we integrate from $ (r_i, \phi_i) $ to $ (r_{min}, \phi_{in}(r_{min})) $. At this point, an outgoing geodesic follows. Now, the outgoing geodesic with initial point $ (r_{min}, \phi_i' \equiv \phi_{in}(r_{min})) $ satisfies
\begin{equation}\label{geo2}
    \phi_{out} (r) = \int_{r_{min}}^{r} \frac{-dr'}{r'^2\sqrt{(1-\frac{b(r')}{r'})(\frac{e^{-2\Phi}}{\beta^2}-\frac{1}{r'^2})}}  + \phi_i',
\end{equation}
and we integrate to $ r_i $.  

\begin{figure*}[hbt!] 
\centering
 \includegraphics[width=0.4\textwidth]{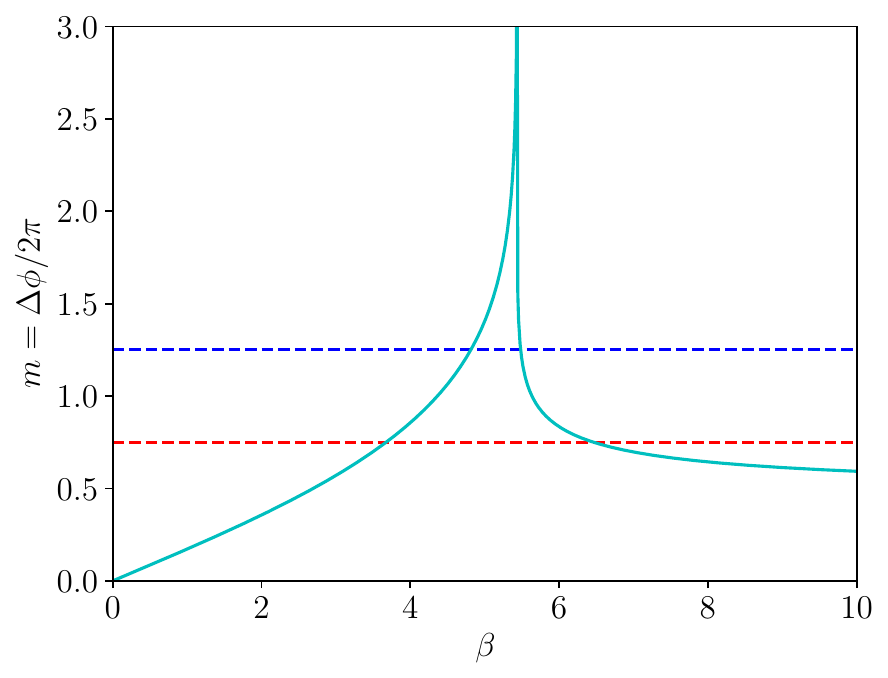}\
\includegraphics[width=0.4\textwidth]{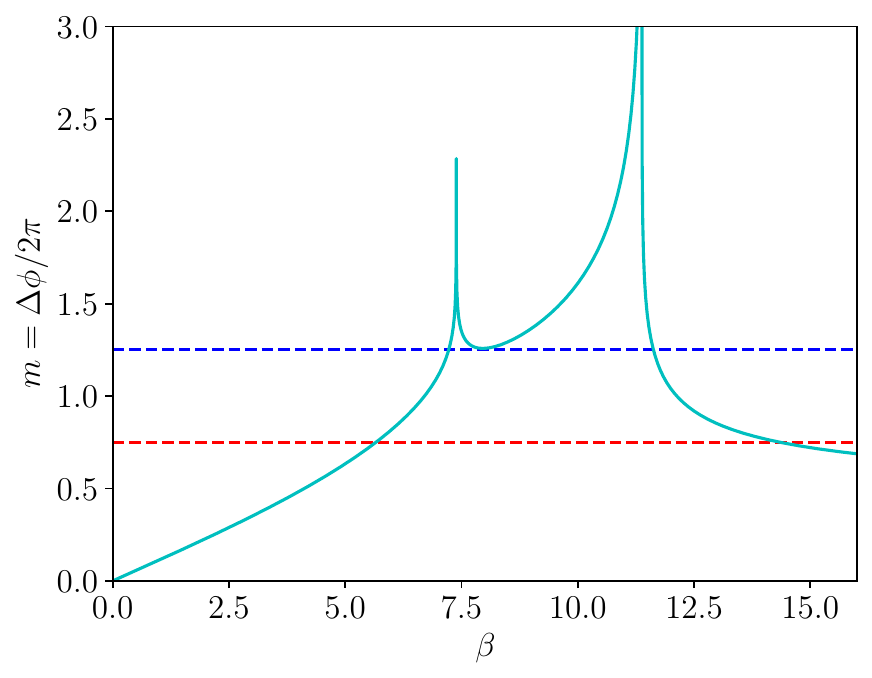}\
\includegraphics[width=0.4\textwidth]{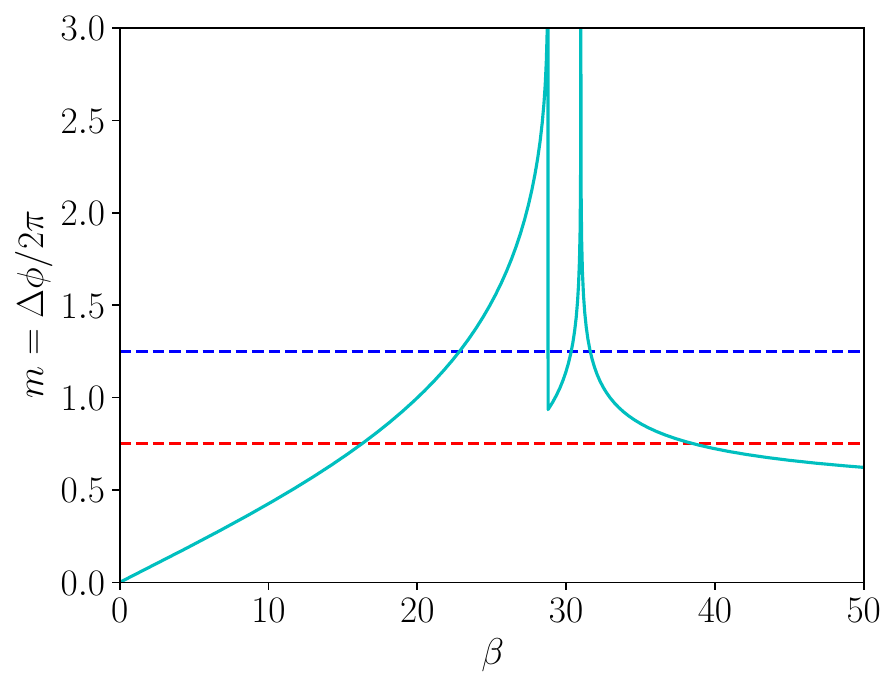}\

\caption{\label{n}
The number of half-orbits $ m = \Delta\phi/2\pi $ as a function of the impact parameter $ \beta $. The red line marks $ m = 3/4 $ and the blue line $ m = 5/4 $.}
\end{figure*}

In figures \ref{rtn1}, \ref{rtn2}, and \ref{rtn3}, the ray-tracing of the wormhole geometry is shown for 
$n = 0, 1, 2$  respectively. The case 
$n = 0$ has two maximum of the potential located outside the throat, one at each distinct universe. Clearly, the photons with impact parameters less than the critical impact parameter are the ones that pass through the throat to the other universe. The case 
$n = 1$  has two maxima of the potential $V_n$ . The first is at the throat $r_0$  and the second at $r_2 = 3 r_0 $ (see Table \ref{table2}). Light orbits make a larger number of turns as they approach the critical impact parameter, an effect that is more pronounced at the outer critical point. This is evidenced in Figure \ref{n} where it can be seen that there is a wider range of impact parameters with more than three turns around the outer critical point. The case 
$n = 2$  has two maxima of $V_n$ outside the throat. It is worth noting that there is a very wide range of impact parameters that pass through the throat of the wormhole. In this case, the number of turns also increases as it approaches the impact parameter. Unlike the case 
$n = 1$ , this effect is more pronounced at the inner critical curve. In general, the three cases have a similar behavior. As $n$  increases, the potential  $V_n$  becomes wider and there is a broader relevant range of impact parameters. It is worth noting that the trajectories become more unstable (two trajectories with very similar initial conditions end up very far apart) as we approach the critical points.

The number of times the light ray orbits the wormhole is determined by the impact parameter $ \beta $. We will classify the orbits according to the number of orbits they complete. This involves dividing the impact parameter space. We introduce the number of half-orbits $ m \equiv \Delta\phi / 2\pi $, where $ \Delta\phi = \phi_{\text{final}} - \phi_{\text{initial}} $, which relates to the number of times the orbit crosses the equatorial plane (where $ \cos \phi = 0 $) as $ k = \lceil 2m \rceil $.  To this end, we will consider three types of trajectories: i) Direct emission where light rays cross the equatorial plane once, $ 1/2 < m < 3/4 $, $ k=1 $. ii) Deflected emission where light rays cross the equatorial plane twice, $ 3/4 < m < 5/4 $, $ k=2 $, and iii) Photon ring emission  where light rays cross the equatorial plane at least three times, $ m > 5/4 $, $ k > 3 $.
It is worth mentioning that, there is a special point where the light orbits infinite times the throat which corresponds to the critical impact parameter, $\beta_{c}$, which is listed in Table \ref{table2}.
\begin{table*}[ht!]
\begin{tabular}{|c|c|c|l|}
\hline
$n$ & Critical point $r_c$          & Impact parameter $\beta_c$ & Interpretation \\ \hline
0    & $r_1 = 2r_0$                 & $5.4366$                   & Single critical curve at $r_1$. \\ \hline
1    & $r_0, r_2 = 3r_0$            & $7.3891, 11.3810$          & Two critical curves: at the throat $r_0$ and at $r_2$. \\ \hline
2     & $r_1 = 2r_0, r_3 = 8r_0$     & $28.7838, 30.9882$         & Two critical curves outside the throat. \\ \hline
\end{tabular}
\caption{Unstable critical points $r_c$ and corresponding impact parameters $\beta_c$ for wormhole models with 
$n = 0, 1, 2$}. \label{table2}
\end{table*}
In Figure \ref{n}, we show the plots of $ \Delta\phi / 2\pi $ as a function of the impact parameter $ \beta $ for 
$ n = 0, 1, 2 $. Note that the number of orbits of the trajectory increases as the impact parameter approaches critical values, where, there are an infinite number of orbits. 

\begin{comment}
Para el caso de $n = 0$ las regiones de $\beta$ a considerar son:

emisión directa: $\beta \in (0,3.8438)\cup(6.0060,10)$

lensed: $\beta \in (3.8438,4.9249) \cup(5.4454,6.0060)$

ligth ring: $\beta \in (4.9249,5.4454)$

Para el caso de $n = 1$ las regiones de $\beta$ a considerar son:

emisión directa: $\beta \in (0,6.0086)\cup(13.4845,16)$

lensed: $\beta \in (6.0086,8.8761) \cup(11.5563,13.4845)$

ligth ring: $\beta \in (8.8761,11.5563)$

Para el caso de $n = 2$ las regiones de $\beta$ a considerar son:

emision directa: $\beta \in (0,16.9169)\cup(35.3353,50)$

lensed: $\beta \in (16.9169,23.3233) \cup(28.7839,30.4983) 
   \cup(31.3813,35.3353)$

ligth ring: $\beta \in (23.3233,28.7839) \cup (30.4983,31.3813)$
\end{comment}

\section{Thin accretion disks.}\label{thin}
To model the emission from a finite disk, we begin with the radiative transport equation, which, neglecting scattering, is \cite{rybicki1991radiative}
\begin{eqnarray} \label{rad}
    \frac{d}{d \lambda} \left(\frac{I_\nu}{\nu^3}\right) = \frac{j_\nu}{\nu^2} - \nu\alpha_\nu \frac{I_\nu}{\nu^3},
\end{eqnarray}
where $ I_\nu $ is the intensity for frequency $ \nu $, $ j_\nu $ is the emissivity, and $ \alpha_\nu $ is the opacity. Analytical models can be implemented by making certain simplifications such as ignoring absorption effects $ \alpha_\nu = 0 $ and assuming the source has monochromatic emission, i.e., $ j_\nu \sim \nu^2 $. For a geometrically thin accretion disk, \eqref{rad} implies that $ I_\nu / \nu^3 $ is conserved along the photon's path. Additionally, we assume isotropic emission $ I^{em}_\nu = I(r) $.

To describe different stages of temporal evolution of an accretion disk, we will consider three toy models of emission (for details, see \cite{Guerrero:2022qkh,Gralla:2019xty}).

Model I. Emission starts at $ r_{ISCO} $ and is modeled as
\begin{equation}\label{em1}
    I_I^{em} = \frac{1}{(r-(r_{ISCO}-1))^2}
\end{equation}
if $ r > r_{ISCO} $, and zero otherwise. The innermost stable circular orbit radius $ r_{ISCO} $ corresponds to the radius where the effective potential of massive particles \eqref{effpot} (with $ \epsilon = 1 $ and $ \theta = \pi/2 $) has an inflection point, i.e., $ V'(r_{ISCO}) = 0 $ and $ V''(r_{ISCO}) = 0 $.

Model II. Emission starts at the outermost unstable critical curve radius $ r_{ph} $ and is modeled as
\begin{equation}\label{em2}
    I_{II}^{em} = \frac{1}{(r-(r_{ph}-1))^3}
\end{equation}
if $ r > r_{ph} $, and zero otherwise.

\begin{figure*}[hbt!]
\centering
 \includegraphics[width=0.3\textwidth]{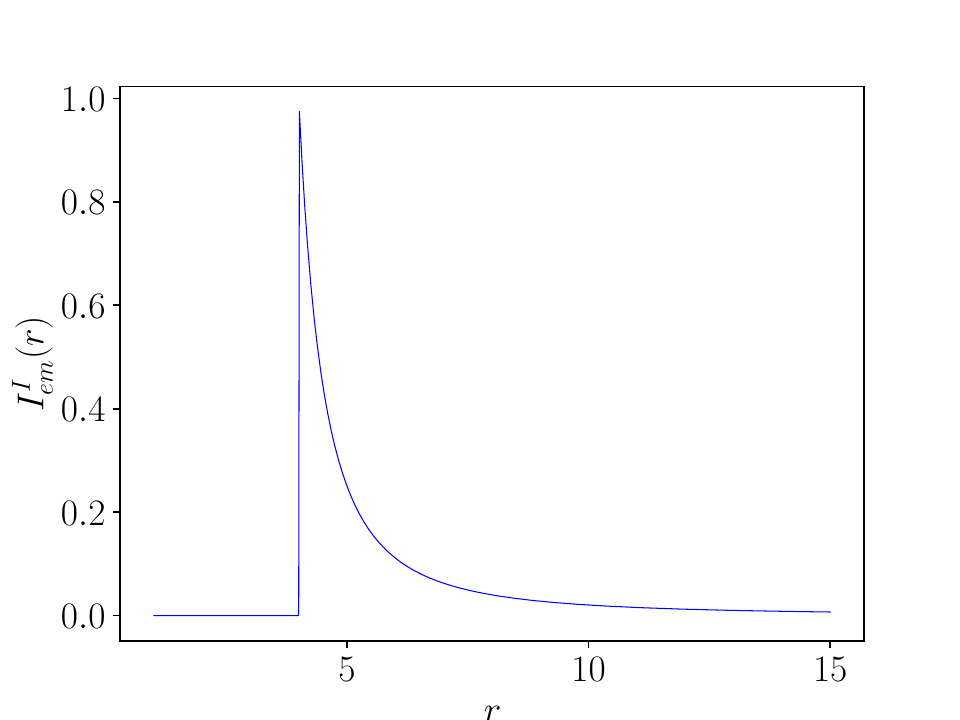}\
\includegraphics[width=0.3\textwidth]{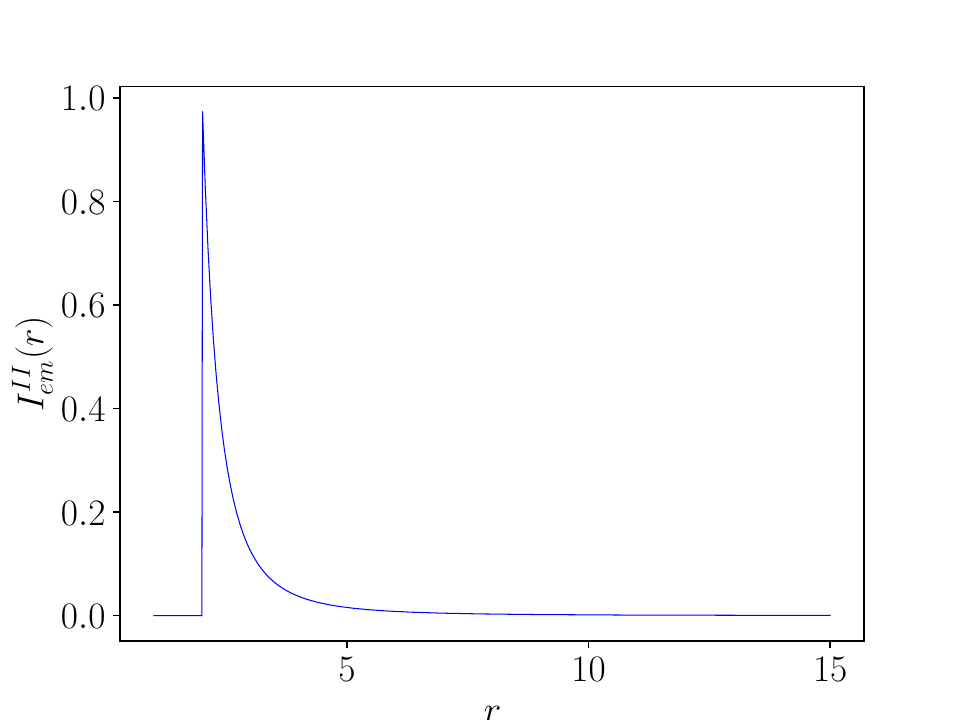}\
\includegraphics[width=0.3\textwidth]{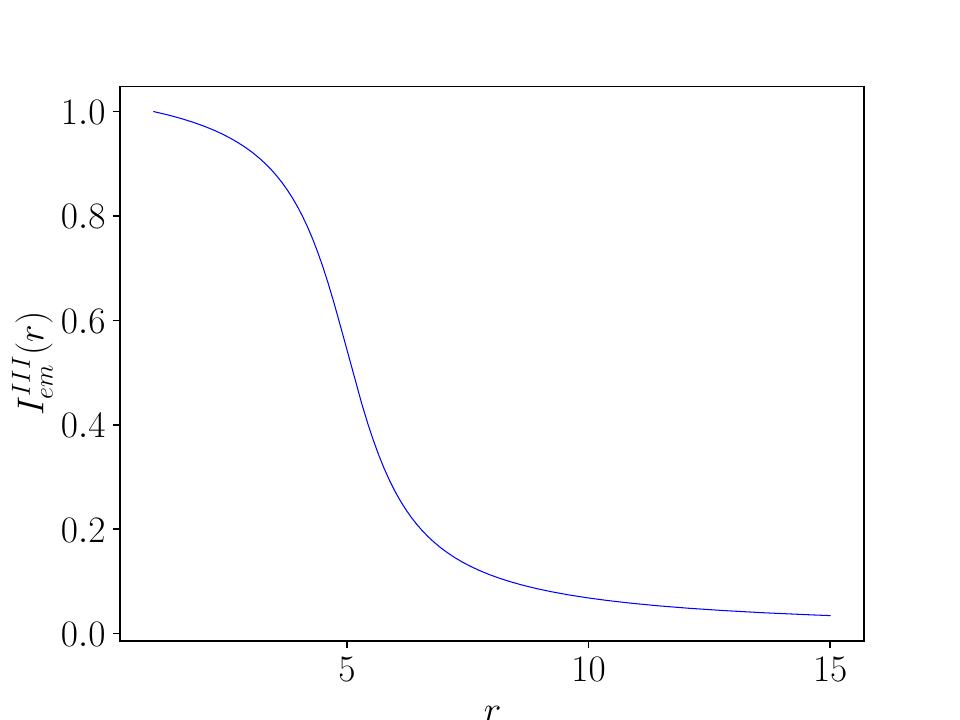}\

\caption{\label{pem1}
Plots of emitted intensity as a function of $ r $ for 
$ n = 0 $}, model $ I $ \eqref{em1} (left), model $ II $ \eqref{em2} (center), model $ III $ \eqref{em3} (right). The plots for $ n = 1 $ and $ n = 2 $ are similar.
\end{figure*}

Model III. Emission starts at the throat $ r_{0} $ and is modeled as
\begin{equation}\label{em3}
    I_{III}^{em} = \frac{\pi/2 - \arctan(r-5)}{\pi/2 - \arctan(r_0-5)}
\end{equation}
if $ r > r_{0} $, and zero otherwise.  Figure \ref{pem1} shows the three emission models for a wormhole with 
$ n = 0 $. For other cases, the profile is similar, with only the position of the peak changing. 

\begin{figure*}[hbt!] 
\centering
 \includegraphics[width=0.3\textwidth]{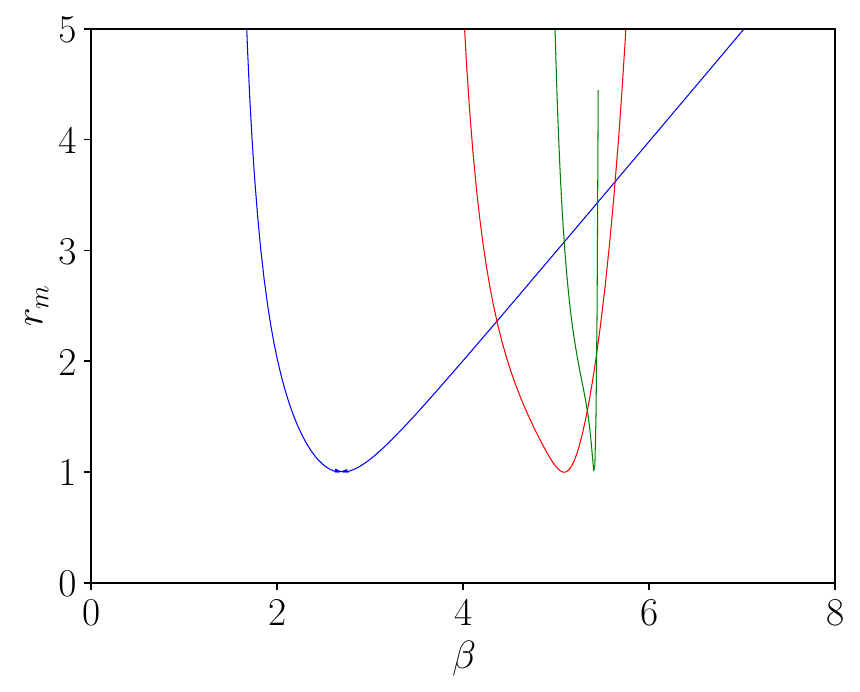}\
\includegraphics[width=0.3\textwidth]{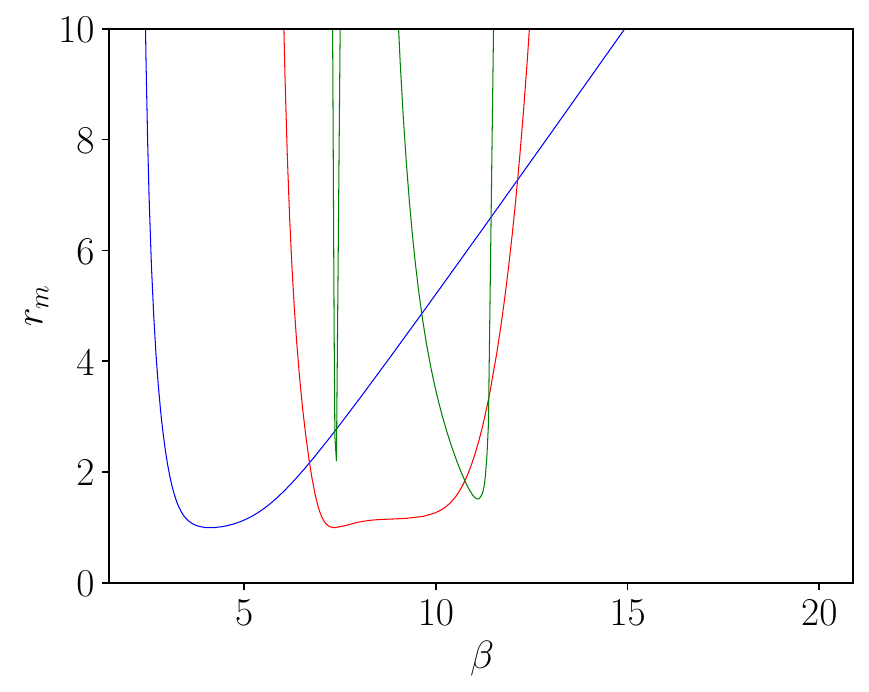}\
\includegraphics[width=0.3\textwidth]{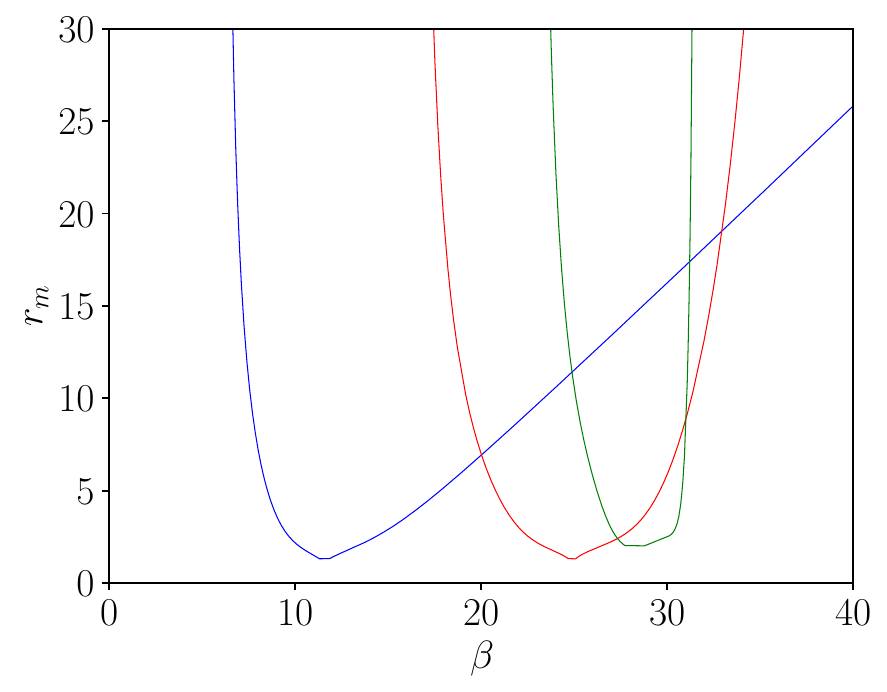}\
\caption{\label{rmb}
Transfer function $ r_m(\beta) $ as a function of the impact parameter $ \beta $. The blue line corresponds to direct emission ($ m=1 $), the red line to deflected emission ($ m=2 $), and the green line to photon ring emission ($ m=3 $). This is shown for cases $ n = 0 $ (left), $ n = 1 $ (center), and $ n = 2 $ (right). }
\end{figure*}

The intensity received by the observer depends on two factors: gravitational redshift and additional intersections of trajectories with the accretion disk contribute to the luminosity. To address the first factor, we start with the frequency in the rest frame of the disk, $ \nu_e $, and its associated intensity $ I_{\nu_e} $. According to the Liouville's theorem, the frequency of the photon measured by a distant observer is $ \nu_o $, with the observed intensity $ I_{\nu_o}^{ob} = (\nu_e/\nu_o)^3 I_{\nu_e}^{em} $. In the spherically symmetric parametrization considered in this work, we have $ I_{\nu_o}^{ob} = e^{4\Phi/3} I_{\nu_e} $. Integrating over all frequencies $ \nu_e $, $ I^{ob} = \int d\nu_e I^{ob}_{\nu_e} $ \cite{Guerrero:2022qkh}, we obtain the result $ I^{ob} = e^{4\Phi} I(r) $. To include the second effect, we need to calculate the contribution from each intersection of the orbits with the accretion disk as follows:
\begin{eqnarray}
    I^{ob} (\beta) = \sum_m e^{4\Phi(r)} I(r)\bigg|_{r_m(\beta)},
\end{eqnarray}
where $ r_m(\beta) $ is the transfer function representing the $ m $-th intersection of the light ray with impact parameter $ \beta $ with the accretion disk. In this work, we consider only $ m = 1, 2, 3 $, corresponding to direct emission, deflected emission, and photon ring emission. Higher-order cases contribute very little to the total luminosity. Figure \ref{rmb} shows the transfer functions as a function of impact parameters for $ n = 0, 1, 2 $. 
In all three cases, there is a direct emission with a very steep slope for small impact parameters, and it is an almost diagonal straight line for larger values. Furhtermore, near the critical impact parameters, the  emission has a steeper slope than the deflected emission. In the cases $ n = 1 $ and $ n = 2 $, the curves for deflected emission and photon ring emission are wider because they have two critical impact parameters. Particularly in the case $ n = 1 $, the curve corresponding to photon ring emission is split into two; the first is particularly narrow and steep. This is because there is a very small range of impact parameters of this type (see \ref{n}). The width of the curves corresponding to deflected and photon ring emission increases with $ n $.

\begin{figure*}[hbt!] 
\centering
 \includegraphics[width=0.3\textwidth]{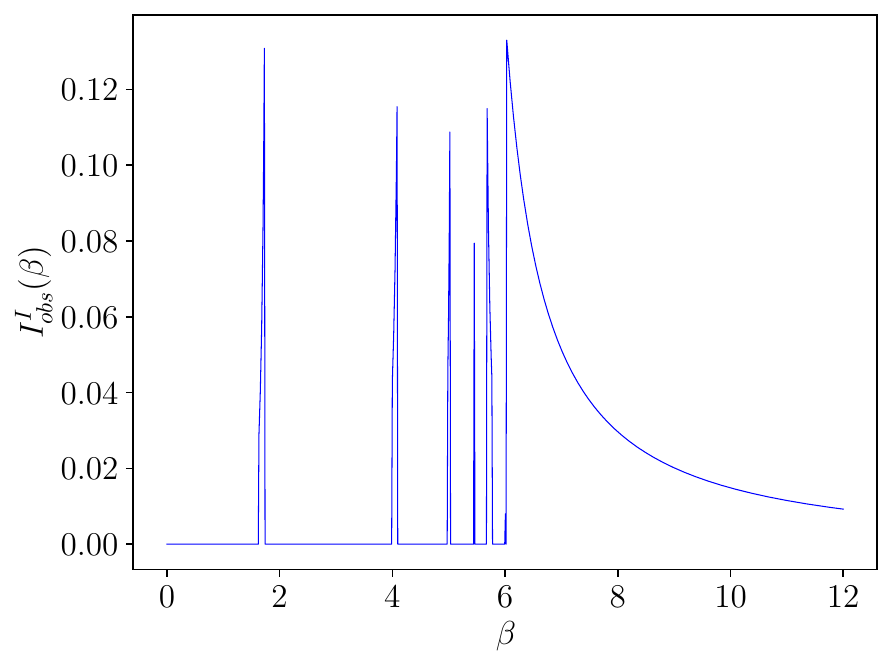}\
\includegraphics[width=0.3\textwidth]{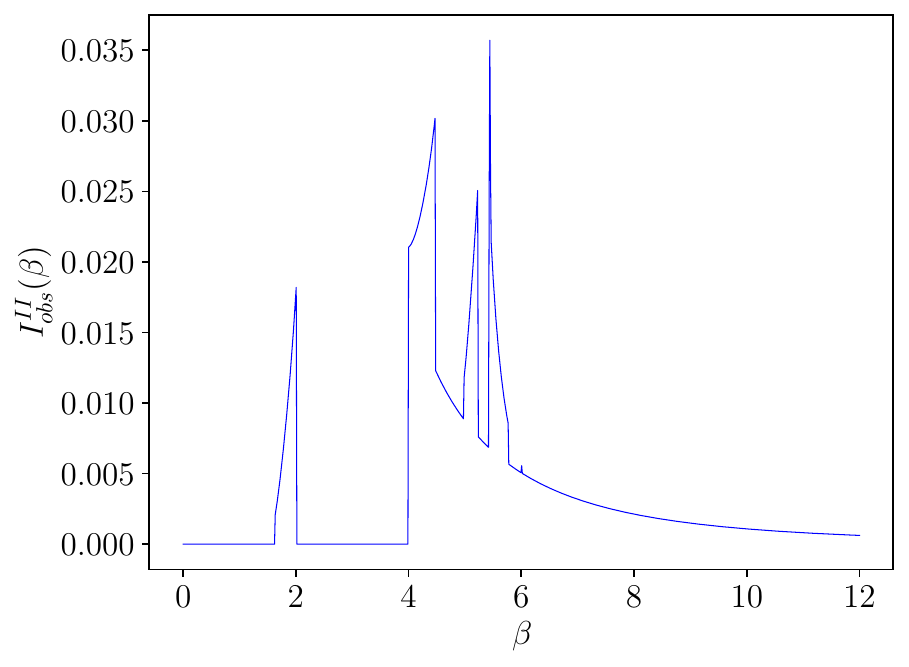}\
\includegraphics[width=0.3\textwidth]{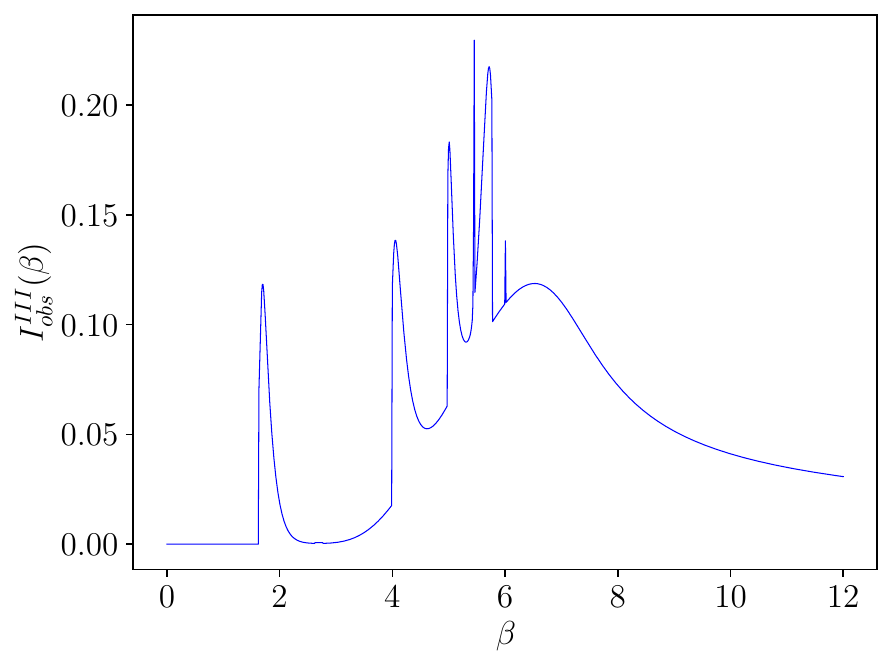}\

 \includegraphics[width=0.3\textwidth]{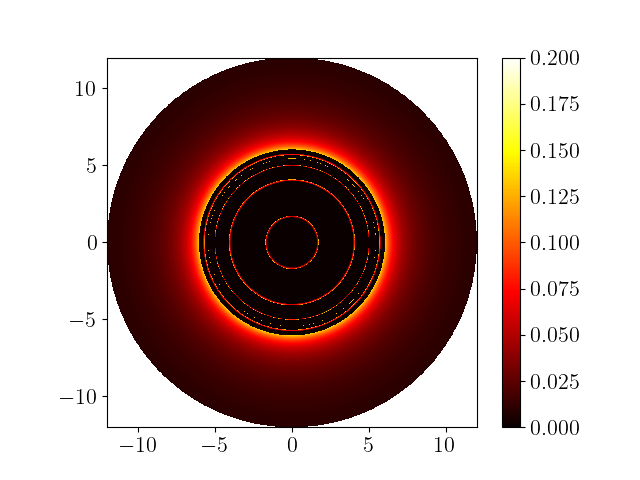}\
\includegraphics[width=0.3\textwidth]{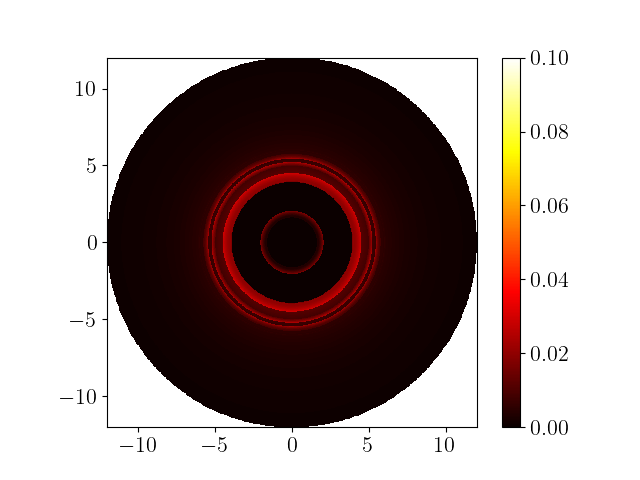}\
\includegraphics[width=0.3\textwidth]{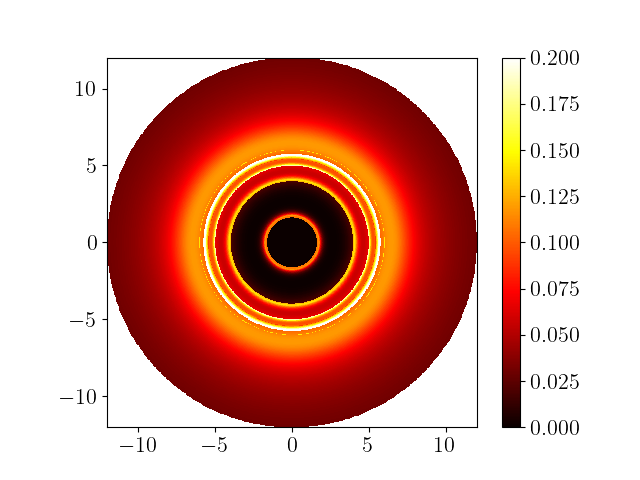}\
\caption{\label{obs1}
Observed intensity as a function of the impact parameter $ \beta $ for $ n = 0 $}, Model $ I $ (left), Model $ II $ (center), Model $ III $ (right).
\end{figure*}

\begin{figure*}[hbt!] 
\centering
 \includegraphics[width=0.3\textwidth]{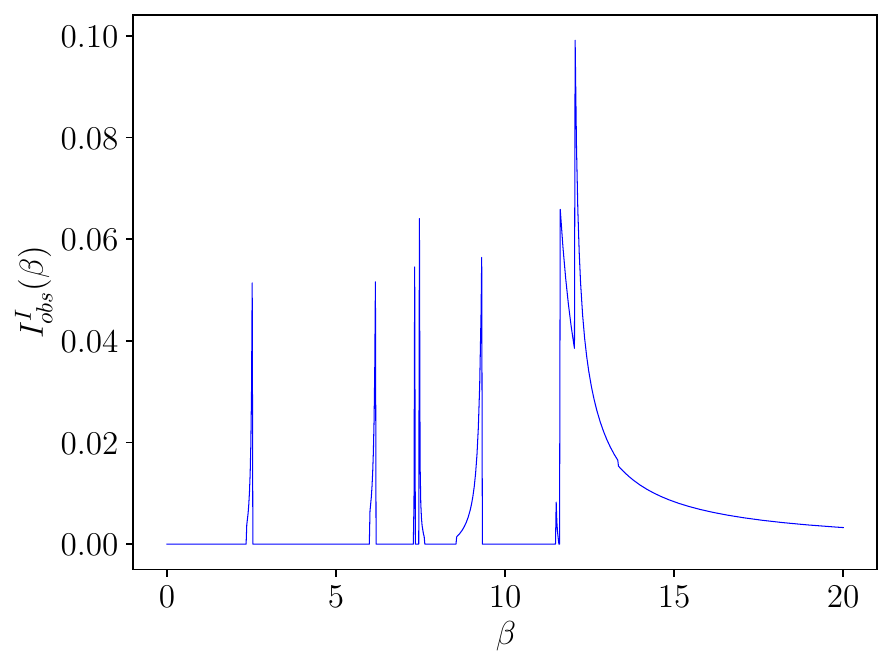}\
\includegraphics[width=0.3\textwidth]{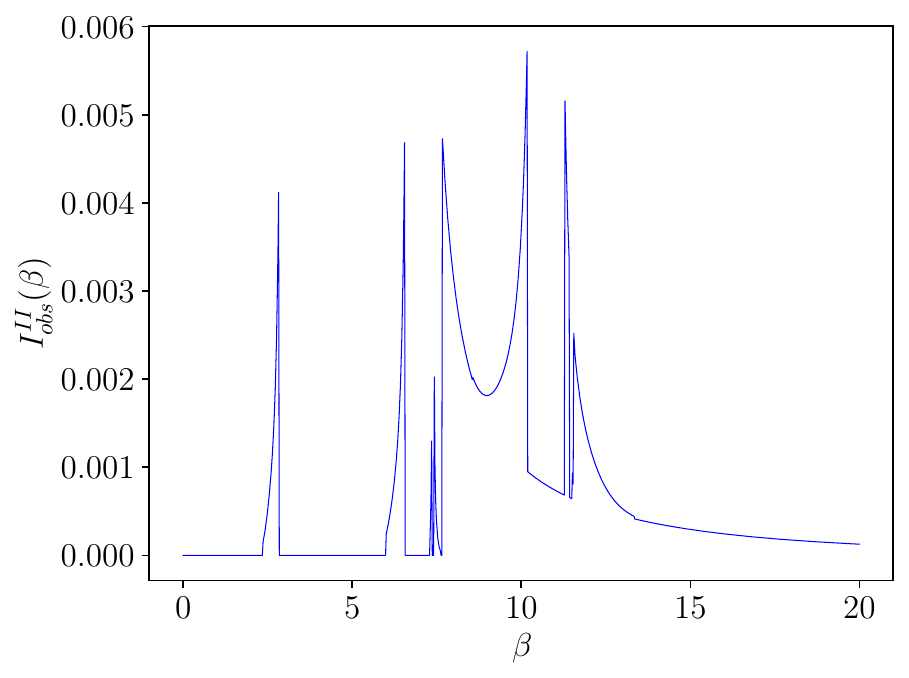}\
\includegraphics[width=0.3\textwidth]{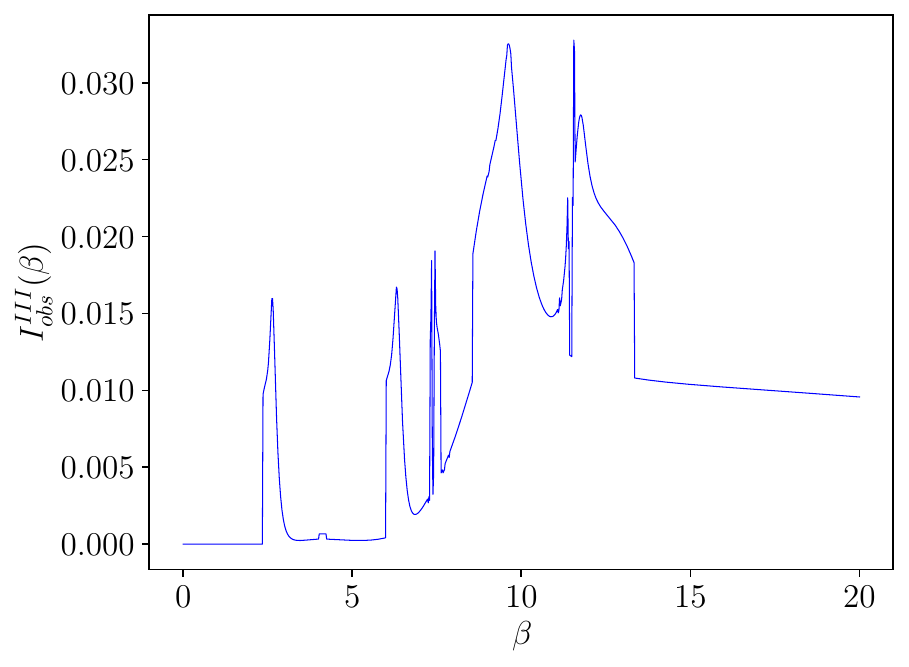}\

 \includegraphics[width=0.3\textwidth]{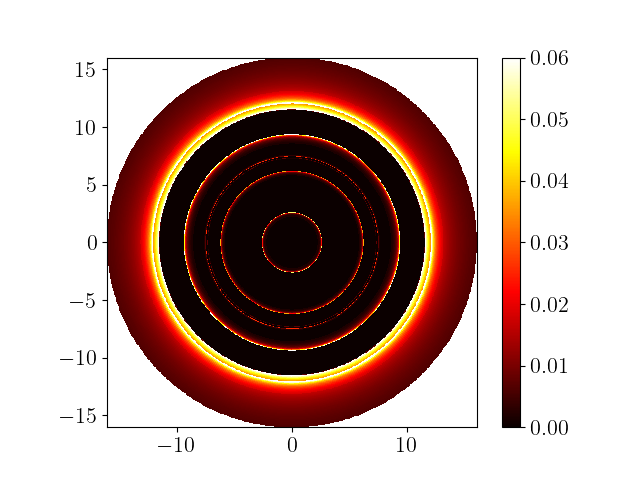}\
\includegraphics[width=0.3\textwidth]{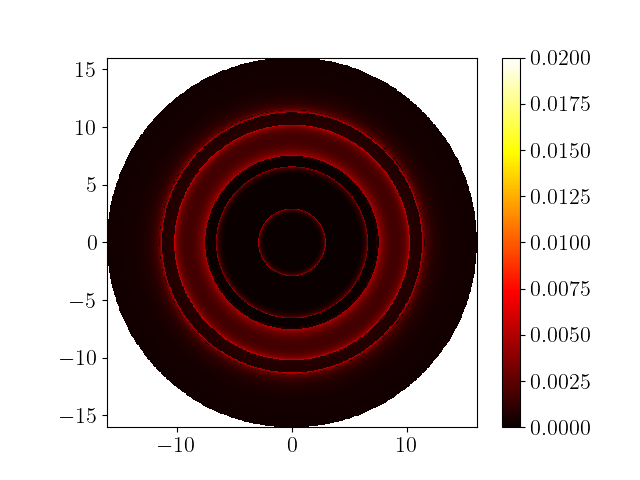}\
\includegraphics[width=0.3\textwidth]{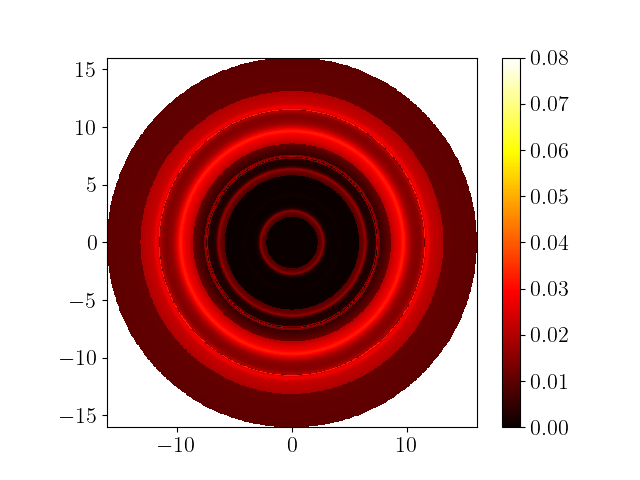}\
\caption{\label{obs2}
Observed intensity as a function of the impact parameter $ \beta $ for $ n = 1 $}, Model $ I $ (left), Model $ II $ (center), Model $ III $ (right).
\end{figure*}

\begin{figure*}[hbt!] 
\centering
 \includegraphics[width=0.3\textwidth]{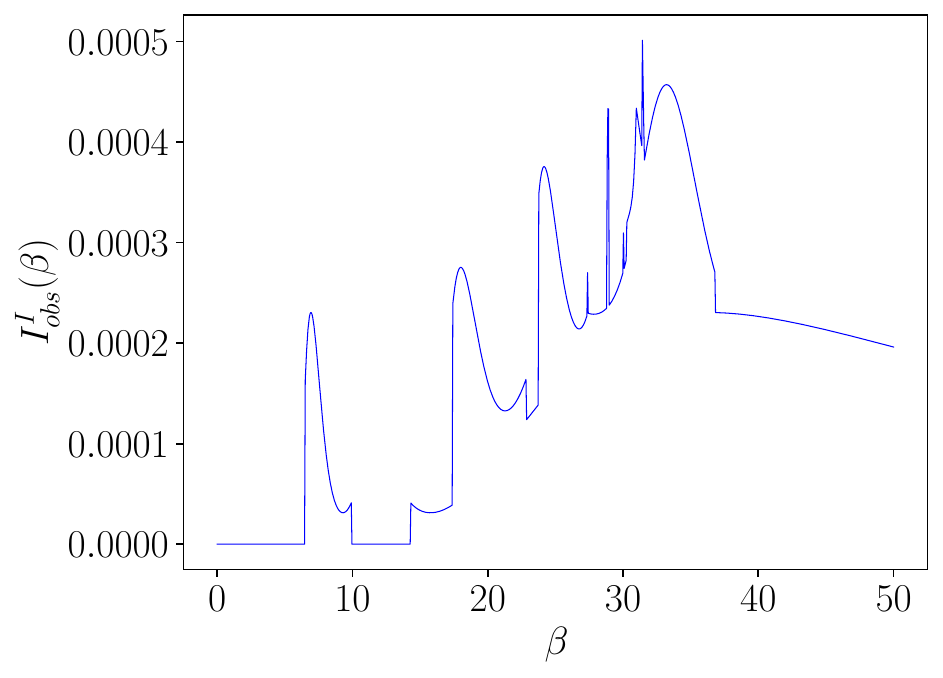}\
\includegraphics[width=0.3\textwidth]{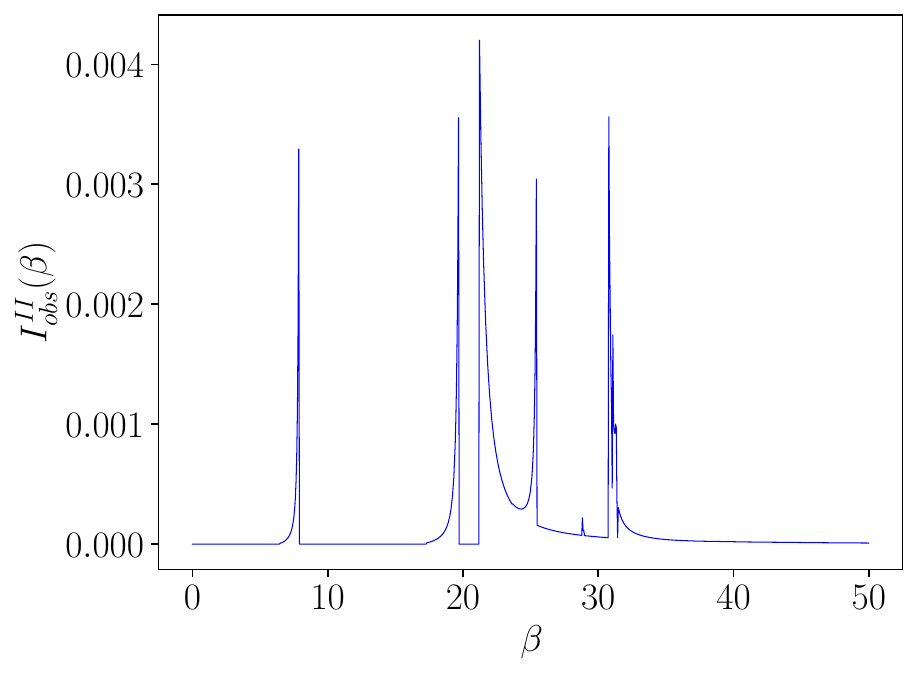}\
\includegraphics[width=0.3\textwidth]{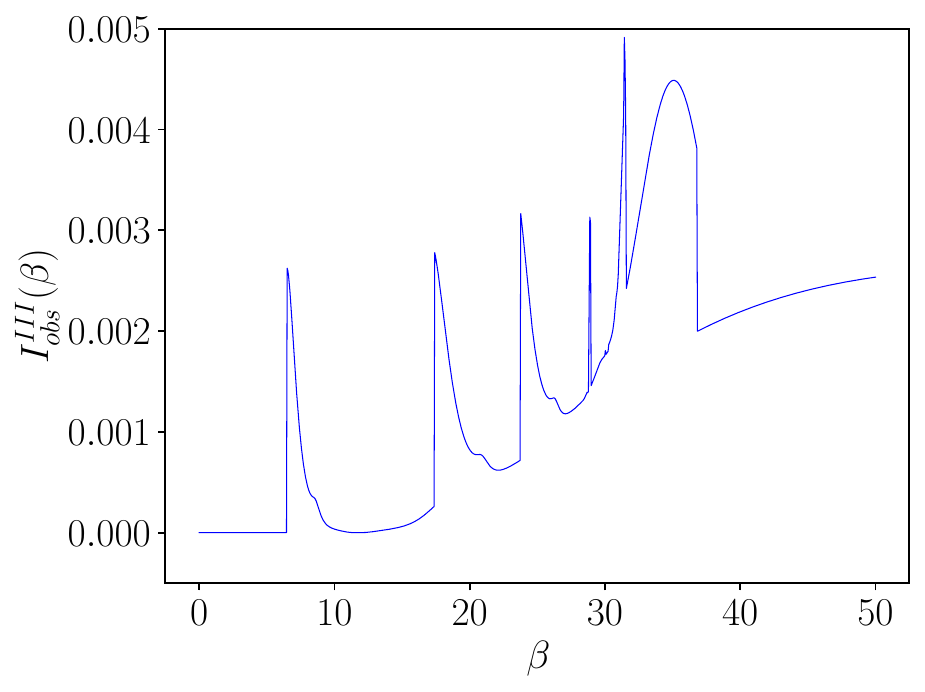}\

 \includegraphics[width=0.3\textwidth]{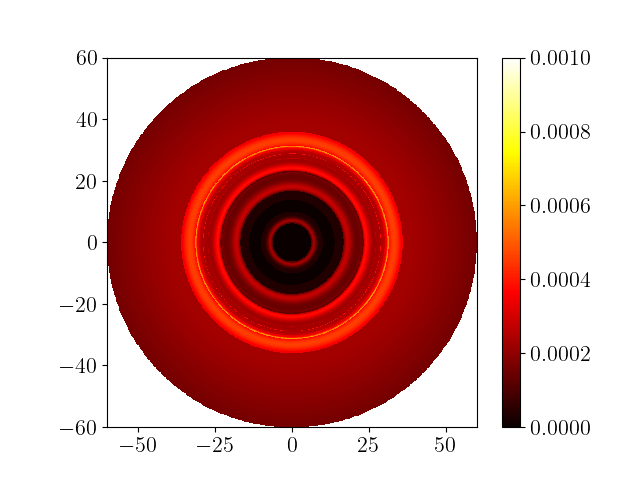}\
\includegraphics[width=0.3\textwidth]{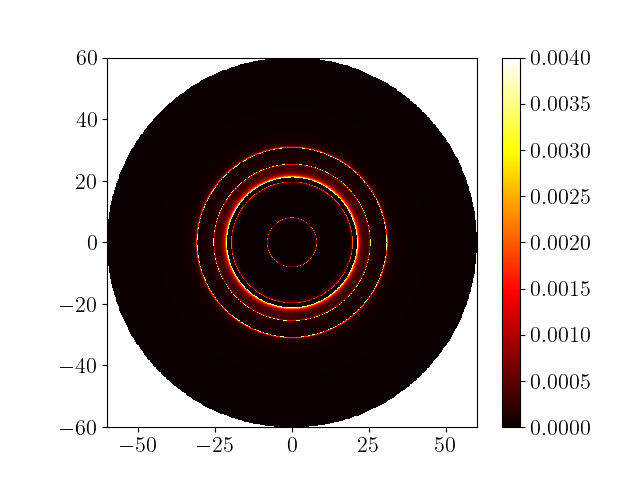}\
\includegraphics[width=0.3\textwidth]{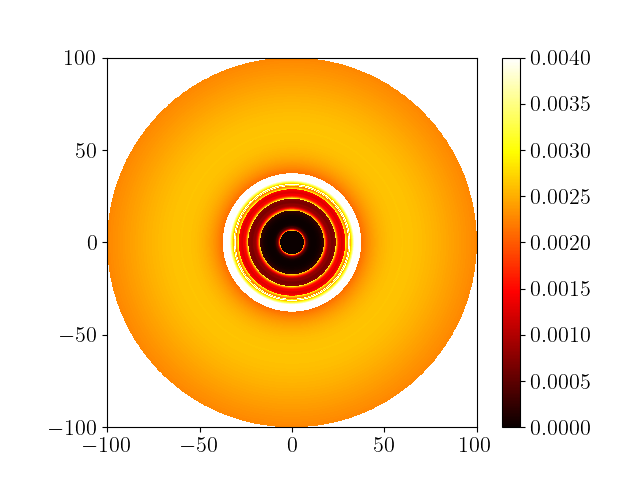}\
\caption{\label{obs3}
Observed intensity as a function of the impact parameter $ \beta $ for $ n = 2 $}, Model $ I $ (left), Model $ II $ (center), Model $ III $ (right). 
\end{figure*}

In figures (\ref{obs1}), (\ref{obs2}), and (\ref{obs3}), the observed intensity and optical appearance are shown. For $ n = 0 $, models I and III have similar intensities, both surpassing model II by two orders of magnitude. Model I exhibits 6 peaks, with the last peak broader than the others. Model II shows three distinct peaks. Model III also displays 6 peaks, with the last peak being wider. For $ n = 1 $, a similar pattern is observed as for $ n = 0 $: models I and III have similar intensities, but this time they are only one order of magnitude higher than those of model II. Model I shows 6 well-defined peaks. Model II has 5 well-defined peaks. Model III loses some definition with 5 peaks, but they are wider. For $ n = 2 $, the intensities of models II and III surpass those of model 1 by one order of magnitude. Model I displays 4 broad peaks and one well-defined peak. Model II has 5 well-defined peaks. Model III again resembles model I with 4 broad peaks and one well-defined peak. Model II decays very rapidly, while models I and II decay much slowly than model II, moreover they never go to zero. 
Model I results in well defined narrow peaks for $n = 0,1$ but wider peaks for $n = 2$. Model III results in wider peaks for all three cases.
Undoubtedly, the number of critical curves plays an important role in the position and intensity of the observed spectrum. These characteristics would not only allow differentiating a traversable wormhole with an effective potential symmetric with respect to its throat from Kerr-type black holes, but also from asymmetric traversable wormholes.

\section{Conclusions}
In this work, we construct traversable wormholes with multiple critical curves. We study in detail the behavior of null geodesics near the peaks of the associated effective potential. In particular, we consider three types of orbits associated with a thin accretion disk: direct emission, deflected emission, and photon ring emission. To analyze the behavior of light rays, we use the ray tracing technique. To study the optical appearance associated with the thin accretion disk, we consider three emission models: starting at the innermost stable circular orbit radius, starting at the furthest critical curves, and starting at the throat. Our results open exciting possibilities for future observational tests using next-generation radiointerferometry instruments, such as the Event Horizon Telescope (EHT) upgrades or the proposed ngEHT (next-generation Event Horizon Telescope). Current observations of black hole shadows, such as those of M87* and Sgr A*, provide angular resolutions on the order of tens of microarcseconds, which are sufficient to detect gross features like the shadow size and asymmetry. However, distinguishing finer details, such as the multiple emission rings predicted by our wormhole models, requires significantly higher resolution and sensitivity. Upcoming advancements in radiointerferometry, including higher frequency observations and larger baseline arrays, could enable the detection of these subtle features. For instance, the differences in the light emission patterns, determined by the number and position of unstable critical curves in our models, might produce distinct signatures in the photon ring structure. Such observations would not only test the predictions of general relativity but also offer a pathway to identify compact objects that deviate from the classical black hole paradigm. This makes our symmetric wormhole solutions with multiple critical curves a promising target for future high-precision imaging experiments

There are some aspects left open for future exploration. One of them is the study of quasinormal modes in this geometry, which would be another unique indicator of the black hole mimicker we are presenting here. Another aspect to study is what optimal values would allow tidal forces to permit traversability through our wormhole solution. As we saw, to ensure the existence of multiple critical curves, we had to introduce a non-zero redshift function that introduces radial tidal forces. Another aspect is to explore other types of shape functions since, as we mentioned in the work, the one we used here was chosen for convenience as numerical calculations can be demanding depending on the profile considered. However, we leave these and other aspects for future work.
\\

\section{acknowledgements}
E. C. is funded by the Beatriz Galindo contract BG23/00163 (Spain). E. C. acknowledge Generalitat Valenciana through PROMETEO PROJECT CIPROM/2022/13.

%\section*{References}
\bibliography{references.bib}

\begin{thebibliography}{10}

\bibitem{LIGOScientific:2016aoc}
B.~P. Abbott et~al.
\newblock {Observation of Gravitational Waves from a Binary Black Hole Merger}.
\newblock {\em Phys. Rev. Lett.}, 116(6):061102, 2016.

\bibitem{LIGOScientific:2016sjg}
B.~P. Abbott et~al.
\newblock {GW151226: Observation of Gravitational Waves from a 22-Solar-Mass
  Binary Black Hole Coalescence}.
\newblock {\em Phys. Rev. Lett.}, 116(24):241103, 2016.

\bibitem{LIGOScientific:2017bnn}
Benjamin~P. Abbott et~al.
\newblock {GW170104: Observation of a 50-Solar-Mass Binary Black Hole
  Coalescence at Redshift 0.2}.
\newblock {\em Phys. Rev. Lett.}, 118(22):221101, 2017.
\newblock [Erratum: Phys.Rev.Lett. 121, 129901 (2018)].

\bibitem{LIGOScientific:2017ycc}
B.~P. Abbott et~al.
\newblock {GW170814: A Three-Detector Observation of Gravitational Waves from a
  Binary Black Hole Coalescence}.
\newblock {\em Phys. Rev. Lett.}, 119(14):141101, 2017.

\bibitem{LIGOScientific:2017vox}
B.~. P.~. Abbott et~al.
\newblock {GW170608: Observation of a 19-solar-mass Binary Black Hole
  Coalescence}.
\newblock {\em Astrophys. J. Lett.}, 851:L35, 2017.

\bibitem{EventHorizonTelescope:2019dse}
Kazunori Akiyama et~al.
\newblock {First M87 Event Horizon Telescope Results. I. The Shadow of the
  Supermassive Black Hole}.
\newblock {\em Astrophys. J. Lett.}, 875:L1, 2019.

\bibitem{EventHorizonTelescope:2019uob}
Kazunori Akiyama et~al.
\newblock {First M87 Event Horizon Telescope Results. II. Array and
  Instrumentation}.
\newblock {\em Astrophys. J. Lett.}, 875(1):L2, 2019.

\bibitem{EventHorizonTelescope:2019jan}
Kazunori Akiyama et~al.
\newblock {First M87 Event Horizon Telescope Results. III. Data Processing and
  Calibration}.
\newblock {\em Astrophys. J. Lett.}, 875(1):L3, 2019.

\bibitem{EventHorizonTelescope:2019ths}
Kazunori Akiyama et~al.
\newblock {First M87 Event Horizon Telescope Results. IV. Imaging the Central
  Supermassive Black Hole}.
\newblock {\em Astrophys. J. Lett.}, 875(1):L4, 2019.

\bibitem{EventHorizonTelescope:2019pgp}
Kazunori Akiyama et~al.
\newblock {First M87 Event Horizon Telescope Results. V. Physical Origin of the
  Asymmetric Ring}.
\newblock {\em Astrophys. J. Lett.}, 875(1):L5, 2019.

\bibitem{EventHorizonTelescope:2019ggy}
Kazunori Akiyama et~al.
\newblock {First M87 Event Horizon Telescope Results. VI. The Shadow and Mass
  of the Central Black Hole}.
\newblock {\em Astrophys. J. Lett.}, 875(1):L6, 2019.

\bibitem{Guerrero:2021pxt}
Merce Guerrero, Gonzalo~J. Olmo, and Diego Rubiera-Garcia.
\newblock {Double shadows of reflection-asymmetric wormholes supported by
  positive energy thin-shells}.
\newblock {\em JCAP}, 04:066, 2021.

\bibitem{Gralla:2019xty}
Samuel~E. Gralla, Daniel~E. Holz, and Robert~M. Wald.
\newblock {Black Hole Shadows, Photon Rings, and Lensing Rings}.
\newblock {\em Phys. Rev. D}, 100(2):024018, 2019.

\bibitem{Morris:1988cz}
M.~S. Morris and K.~S. Thorne.
\newblock {Wormholes in space-time and their use for interstellar travel: A
  tool for teaching general relativity}.
\newblock {\em Am. J. Phys.}, 56:395--412, 1988.

\bibitem{Morris:1988tu}
M.~S. Morris, K.~S. Thorne, and U.~Yurtsever.
\newblock {Wormholes, Time Machines, and the Weak Energy Condition}.
\newblock {\em Phys. Rev. Lett.}, 61:1446--1449, 1988.

\bibitem{Visser:1995cc}
Matt Visser.
\newblock {\em {Lorentzian wormholes: From Einstein to Hawking}}.
\newblock 1995.

\bibitem{Visser:2003yf}
Matt Visser, Sayan Kar, and Naresh Dadhich.
\newblock {Traversable wormholes with arbitrarily small energy condition
  violations}.
\newblock {\em Phys. Rev. Lett.}, 90:201102, 2003.

\bibitem{Alcubierre:2017pqm}
Miguel Alcubierre.
\newblock {\em {Wormholes, Warp Drives and Energy Conditions}}, volume 189.
\newblock Springer, 2017.

\bibitem{Lobo:2005us}
Francisco S.~N. Lobo.
\newblock {Phantom energy traversable wormholes}.
\newblock {\em Phys. Rev. D}, 71:084011, 2005.

\bibitem{Blazquez-Salcedo:2021udn}
Jose~Luis Bl\'azquez-Salcedo, Christian Knoll, and E.~Radu.
\newblock {Einstein-Dirac-Maxwell wormholes: ansatz, construction and
  properties of symmetric solutions}.
\newblock 8 2021.

\bibitem{Bambi:2021qfo}
Cosimo Bambi and Dejan Stojkovic.
\newblock {Astrophysical Wormholes}.
\newblock {\em Universe}, 7(5):136, 2021.

\bibitem{Capozziello:2020zbx}
Salvatore Capozziello, Orlando Luongo, and Lorenza Mauro.
\newblock {Traversable wormholes with vanishing sound speed in $f(R)$ gravity}.
\newblock {\em Eur. Phys. J. Plus}, 136(2):167, 2021.

\bibitem{Blazquez-Salcedo:2020czn}
Jose~Luis Bl\'azquez-Salcedo, Christian Knoll, and Eugen Radu.
\newblock {Traversable wormholes in Einstein-Dirac-Maxwell theory}.
\newblock {\em Phys. Rev. Lett.}, 126(10):101102, 2021.

\bibitem{Berry:2020tky}
Thomas Berry, Francisco S.~N. Lobo, Alex Simpson, and Matt Visser.
\newblock {Thin-shell traversable wormhole crafted from a regular black hole
  with asymptotically Minkowski core}.
\newblock {\em Phys. Rev. D}, 102(6):064054, 2020.

\bibitem{Maldacena:2020sxe}
Juan Maldacena and Alexey Milekhin.
\newblock {Humanly traversable wormholes}.
\newblock {\em Phys. Rev. D}, 103(6):066007, 2021.

\bibitem{Wielgus:2020uqz}
Maciek Wielgus, Jiri Horak, Frederic Vincent, and Marek Abramowicz.
\newblock {Reflection-asymmetric wormholes and their double shadows}.
\newblock {\em Phys. Rev. D}, 102(8):084044, 2020.

\bibitem{Blazquez-Salcedo:2020nsa}
Jose~Luis Bl\'azquez-Salcedo, Xiao~Yan Chew, Jutta Kunz, and Dong-Han Yeom.
\newblock {Ellis wormholes in anti-de Sitter space}.
\newblock {\em Eur. Phys. J. C}, 81(9):858, 2021.

\bibitem{Xavier:2024iwr}
S\'ergio V. M. C.~B. Xavier, Carlos A.~R. Herdeiro, and Lu\'\i{}s C.~B.
  Crispino.
\newblock {Traversable wormholes and light rings}.
\newblock {\em Phys. Rev. D}, 109(12):124065, 2024.

\bibitem{Konoplya:2021hsm}
R.~A. Konoplya and A.~Zhidenko.
\newblock {Traversable Wormholes in General Relativity}.
\newblock {\em Phys. Rev. Lett.}, 128(9):091104, 2022.

\bibitem{rybicki1991radiative}
G.B. Rybicki and A.P. Lightman.
\newblock {\em Radiative Processes in Astrophysics}.
\newblock A Wiley-Interscience publication. Wiley, 1991.

\bibitem{Guerrero:2022qkh}
Merce Guerrero, Gonzalo~J. Olmo, Diego Rubiera-Garcia, and Diego G\'omez
  S\'aez-Chill\'on.
\newblock {Light ring images of double photon spheres in black hole and
  wormhole spacetimes}.
\newblock {\em Phys. Rev. D}, 105(8):084057, 2022.

\end{thebibliography}
\bibliographystyle{unsrt}
\end{document}